\documentclass[12pt,a4paper]{article}
 \usepackage{amssymb}
 \usepackage{amsmath,epsfig}
 \usepackage{amsfonts,latexsym}
 \usepackage{bbm}
 \usepackage{cite}

 \font\mybb=msbm10 at 12pt
 \def\bb#1{\hbox{\mybb#1}}

\begin{document}

 \rightline{\tt hep-th/0406038}
 \rightline{KCL-MTH-04-07}
 \rightline{SPIN-04/08}
 \rightline{ITP-04/14}
 \rightline{UG-04/02}

 \thispagestyle{empty}
 \vskip 1.3truecm

 \centerline{\LARGE \bf Non-extremal D-instantons}
 \vskip 1truecm

 \centerline{\bf E.~Bergshoeff${}^{1}$, A.~Collinucci${}^{1}$,
U.~Gran${}^{2}$,
 D.~Roest${}^{1}$ and S.~Vandoren${}^{3}$}
 \bigskip
 \centerline{${}^{1}$\ Centre for Theoretical Physics, University of
Groningen}
 \centerline{Nijenborgh 4, 9747 AG Groningen, The Netherlands}
 \centerline{E-mail: {\tt (e.a.bergshoeff, a.collinucci,
d.roest)@phys.rug.nl}}

 \vskip .2truecm
 \centerline{${}^{2}$\ Department of Mathematics, King's College London}
 \centerline{Strand, London WC2R 2LS, United Kingdom }
 \centerline{E-mail: {\tt ugran@mth.kcl.ac.uk}}

 \vskip .2truecm \centerline{${}^{3}$\ Institute for Theoretical Physics,
Utrecht University}
 \centerline{Leuvenlaan 4, 3508 TD Utrecht, The Netherlands}
 \centerline{E-mail: {\tt s.vandoren@phys.uu.nl}}

 \vskip 1.5truecm

\centerline{ABSTRACT} \bigskip

We construct the most general non-extremal deformation of the
D-instanton solution with maximal rotational symmetry. The general
non-supersymmetric solution carries electric charges of the
$SL(2,\mathbb{R})$ symmetry, which correspond to each of the
three conjugacy classes of $SL(2,\mathbb{R})$. Our calculations
naturally generalise to arbitrary dimensions and arbitrary dilaton
couplings.

We show that for specific values of the dilaton coupling parameter, the non-extremal
instanton solutions can be viewed as wormholes of non-extremal Reissner-Nordstr\"om black
holes in one higher dimension. We extend this result by showing that for other
values of the dilaton coupling parameter, the non-extremal instanton
solutions can be uplifted to non-extremal non-dilatonic $p$-branes in $p+1$ dimensions
higher.

Finally, we attempt to consider the solutions as instantons of
(compactified) type IIB superstring theory. In particular, we
derive an elegant formula for the instanton action. We conjecture
that the non-extremal D-instantons can contribute to the
$R^8$-terms in the type IIB string effective action.

\vskip .8truecm

PACS number: 04.50.+h 04.65.+e 11.25.-w

\vfill \eject

\section{Introduction}

Gravity coupled to the two scalars (dilaton and axion) that parameterise an
$SL(2,\bb{R})/SO(2)$ coset space is an important subsector of the low-energy limit of
type IIB superstring theory. Among the different solutions of this system are seven-brane
solutions that carry magnetic charges with respect to the three generators of
$SL(2,\bb{R})$. These magnetic charges combine into a traceless 2 x 2 charge matrix $Q_M$
which transforms in the adjoint representation of $SL(2,\bb{R})$. The combination
$\det(Q_M)$, being invariant under these transformations, labels the three different
conjugacy classes of $SL(2,\bb{R})$. Each pair of solutions in the same conjugacy class
is related via $SL(2,\bb{R})$. On the other hand, two solutions that belong to two
different conjugacy classes can not be related via $SL(2,\bb{R})$.

The three classes of inequivalent seven-brane solutions, whose magnetic charges
correspond to the three conjugacy classes of $SL(2,\bb{R})$, have been constructed in
\cite{Bergshoeff:2002mb}. All of these seven-brane solutions are half-supersymmetric. The
conjugacy class with $\det(Q_M)=0$ has $\mathbb{R}$ isometry in the transverse space and
is represented by the so-called ``circular'' 1/2 BPS D7--brane \cite{Bergshoeff:1996ui}.
The other two classes, with $\det(Q_M)>0$ and $\det(Q_M)<0$, describe seven-branes with
$SO(2)$ and $SO(1,1)$ isometry in the transverse directions, respectively. Actually,
there exist two solutions describing seven-branes with $SO(2)$ isometries
\cite{Bergshoeff:2002mb} whose interpretation has become clear only recently
\cite{Bergshoeff:2004nq}. The first solution describes a set of (positively and
negatively charged) D7-branes which are distributed along a finite line-element in one of
the two transverse directions. A special feature is that the total D7-brane charge
vanishes. The second solution can be obtained by taking the first solution in the limit
of a zero-size line-element. Since the total D7-brane charge vanishes, one is not left
with a single D7-brane but, instead, one obtains a conical space-time with a specific
deficit angle \cite{Bergshoeff:2004nq}.

It is well-known that the electric-magnetic dual of the D7--brane is the
D-instanton\footnote{Sometimes an instanton is called a ``-1-- brane'' due to the fact
that the transverse space fills the complete target space. The D-instanton can be seen as
the $p=-1$ end of the T-duality chain of D$p$-branes.} \cite{Gibbons:1996vg,
Green:1997tv}. The D-instanton is a half-supersymmetric solution of the Euclidean
gravity-dilaton-axion system, where the dilaton and axion parameterise an
$SL(2,\bb{R})/SO(1,1)$ coset, and carries electric charge with respect to the Euclidean
$SL(2,\bb{R})$ symmetry. In complete analogy to the case of seven-branes, the three
Euclidean $SL(2,\bb{R})$ charges combine into a 2 x 2 charge matrix $Q_E$ that transforms
in the adjoint representation of the Euclidean $SL(2,\mathbb{R})$. The D-instanton is
represented by the same conjugacy class that represents the circular D7--brane, i.e.~the
one with $\det(Q_E)=0$.

It is natural to ask, in analogy to the case of seven-branes discussed above, whether
there exist ``exotic'' instantons with electric $SL(2,\bb{R})$ charges corresponding to
the other two conjugacy classes, i.e.~the ones with $\det(Q_E)>0$ and $\det(Q_E)<0$. It
is the aim of this paper to construct and investigate such solutions. For earlier work on
generalised D-instanton solutions, see \cite{Giddings:1989bq, Coule:1989xu, Tseytlin:1997ne,
Kim:1997hq, Einhorn:2000ct, Einhorn:2002am, Gutperle:2002km, Kim:2003js, Khoze:2003yk}. To be
specific, in this paper we will construct solutions with maximal rotational symmetry for
general values of the charge matrix $Q_E$, thus generalising the D--instanton, and for
arbitrary dimension $D$ and dilaton coupling $b$, thus including compactifications of
type IIB string theory. We will always refer to the instanton solutions in the
$\det(Q_E)=0$ conjugacy class  as D-instantons, though we allow for the generalisation to
arbitrary dimension and dilaton coupling.

In contrast to the D-instanton we find that solutions corresponding to the other two
conjugacy classes, in Einstein frame, do not have a flat metric but rather a {\sl
conformally flat} metric (which is implied by the rotational $SO(D)$ symmetry). Moreover,
the scalars can be expressed in terms of one rotationally symmetric function which is
harmonic over the conformally flat space. Unlike the case of seven-branes, where all
three conjugacy classes preserve half supersymmetry \cite{Bergshoeff:2002mb}, we find
that the generalised instantons of the other two conjugacy classes do not preserve any
supersymmetry. It has been noted that the standard D-instanton has a manifest wormhole
geometry in the string frame metric \cite{Gibbons:1996vg, Bergshoeff:1998ry}. Curiously,
we find that this holds for all values of $b$ and $D$ for the $\det Q_E =0$ solution in
string frame and the $\det Q_E > 0$ solution in Einstein frame. In addition, for a
particular value of the dilaton coupling parameter, the same applies to the other
conjugacy class with $\det Q_E < 0$ provided we use the so-called dual frame metric.

One can view the generalised instantons as non-extremal deformations of the
half-super\-symmetric instanton. This point of view is confirmed by the fact that a
subclass of these solutions can be viewed as describing wormholes corresponding to
non-extremal Reissner-Nordstr\"om black holes in one dimension higher. More generally,
for specific values of the dilaton coupling parameter, the non-extremal instanton
solutions can be uplifted to regular non-extremal non-dilatonic $p$-branes in $p+1$
dimensions higher.

Alternatively, we will describe in this paper an attempt to
consider the non-extremal D-instantons as true (albeit
non-supersymmetric) instantons of type IIB superstring theory. In
particular, we will derive an elegant expression for the instanton
action. We conjecture that, whereas the extremal D-instantons
contribute to the $R^4$ terms in the type IIB string effective
action \cite{Green:1997tv}, the non-extremal D-instantons may
contribute to the $R^8$-terms in the same effective action.

This paper is organised as follows. In section \ref{bulk} we discuss the realisation of
the $SL(2,\mathbb{R})$-duality group for the Euclidean case. In section \ref{solution} we
give the generalised instanton solutions mentioned above. At this point we only construct
the bulk solutions without taking care of boundary terms and/or boundary conditions.
Next, in section \ref{wormhole} we discuss the relation to wormholes corresponding to
non-extremal Reissner-Nordstr\"om black holes in one dimension higher. In section
\ref{p-branes} we consider generalisations that uplift to non-extremal $p$-branes in
$p+1$ dimensions higher. The application as true instantons of type IIB string theory
will be investigated in section \ref{positive}. Finally, we discuss our results in
section \ref{conclusions}.

\section{$SL(2,\mathbb{R})$-symmetry}\label{bulk}

In order to discuss the $SL(2,\bb{R})$-duality symmetry for both
Minkowskian and Euclidean signatures of space(-time), it is
convenient to first consider a complexification of all fields and,
next, consider the Minkowski and Euclidean cases as different real
slices, see e.g.~\cite{Bergshoeff:2000qu}.

We thus consider the complexification of $D$-dimensional gravity
coupled to complex scalars that parameterise the coset
\begin{equation}
\frac{SL(2,\bb{C})}{SO(2) \times  SO(1,1)}
\end{equation}
via the scalar matrix\footnote{We will use the following notation:
subscripts ${}_C$ refer to quantities in the complex case. For the
two real slices that we consider we use ${}_M$ and ${}_E$ which
will correspond to Minkowskian and Euclidean signatures of the
space-times, respectively.}
\begin{align}
  {\mathcal{M}}_C = e^{{b \phi /2}} \left(
  \begin{array}{cc} \tfrac{1}{4} b^2 \chi^2+e^{-b \phi} & { \tfrac{1}{2} b
\chi} \\ {\tfrac{1}{2} b \chi} & 1 \end{array}
  \right) \,,
\label{complexscalars}
\end{align}
where $\phi$ is the (complex) dilaton and $\chi$ the (complex)
axion. The constant $b$ parameterises the coupling of the dilaton
to the axion. The corresponding (complex) Lagrangian is given by
\begin{align}
  \mathcal{L}_C
  = \tfrac{1}{2} \sqrt{|g|} \,
    [ {R} + b^{-2} \text{Tr}(\partial \mathcal{M}_C \, \partial
\mathcal{M}^{-1}_C) ]
  = \tfrac{1}{2} \sqrt{|g|} \,
    [ {R} -\tfrac{1}{2} (\partial {\phi})^2
    -\tfrac{1}{2} e^{b {\phi}} (\partial {\chi})^2 ] \, ,
\label{10DIIB}
\end{align}
where the metric $g_{\mu \nu}$ is complex and $b$ is an arbitrary
dilaton coupling parameter. In this case there is an
$SL(2,\bb{C})$ symmetry which acts on the dilaton and axion in the
following way\footnote{The dilaton coupling parameter $b$ in
\eqref{10DIIB} is of course different from the parameter $b$ in
the $SL(2,\mathbb{R})$ transformations below. It should always be
clear from the context what is meant.} :
\begin{equation}
  \mathcal{{M}}_C \rightarrow \Omega_C \, \mathcal{{M}}_C \, \Omega_C^T
  ~~~~ \text{with} ~~~
  \Omega_C = \left( \begin{array}{cc} a & b \\ c & d \end{array}
  \right) \in SL(2,\bb{C}) \,.
\label{SL2C}
\end{equation}
The Einstein frame metric is $SL(2,\bb{C})$-invariant.

We now make two different truncations of this complex system
leading to real fields and real Lagrangians. One choice is to take
\begin{equation}\label{slice}
g^*_{\mu\nu} = g_{\mu\nu}\, , \hskip .5truecm \chi^* = \chi\, ,
\hskip .5truecm \phi^* = \phi\, ,
\end{equation}
and to take the (real) metric to be Minkowskian. For the two
(real) scalars this leads to the coset
\begin{equation}
\frac{SL(2,\bb{R})}{SO(2)}
\end{equation}
 parameterised by $\mathcal{M}_M =
\mathcal{M}_C$ as given in \eqref{complexscalars} with $\phi$ and
$\chi$ both real. The Lagrangian $\mathcal{L}_M$ for this case is
given by \eqref{10DIIB} where both the metric and the two scalars
are real.  The $SL(2,\bb{R})$ symmetry is given by
\begin{equation}
  \mathcal{{M}}_M \rightarrow \Omega_M \, \mathcal{{M}}_M \, \Omega_M^T
  ~~~~ \text{with} ~~~
  \Omega_M = \left( \begin{array}{cc} a&b\\c&d \end{array} \right) \in
SL(2,\bb{R})\, . \label{SL2M}
\end{equation}
Equivalently, the $SL(2,\bb{R})$ transformations can also be
defined as modular transformations on the complex field
\begin{equation}
\tau = \frac{b}{2}\, \chi + i\, e^{-b\phi/2}\ .\label{tau}
\end{equation}
The Lagrangian for the scalars can then be rewritten as\footnote{
Throughout this paper we assume that $b \ne 0$. Note that for
$b=0$ the Euclidean $SL(2,\mathbb{R})$ symmetry degenerates to a
$ISO(1,1)$ symmetry.}
\begin{equation}
\mathcal{L}_M^{scal} =-(2/b^2)|\partial{\tau}/{\rm Im}\,\tau|^2\,
,
\end{equation}
with symmetry
\begin{equation}
\tau \rightarrow \frac{a\tau +b}{c\tau +d}\ ,\qquad  ad-bc=1\ .
\end{equation}
This theory occurs for example as the scalar section of IIB supergravity in $D=10$
Minkowski space-time with dilaton-coupling parameter $b=2$. Other values of $b$ can arise
when considering (truncations of) compactifications of IIB supergravity. For instance, in
$D=3$ one has supersymmetry for $b=2, b= \sqrt {2}, b=\sqrt {4/3}$ and $b=1$.

In the second case, on which we will concentrate in this paper, we
first redefine $\chi \rightarrow i\chi$ and next impose the same
reality conditions \eqref{slice} with the only difference that we
now take the (real) metric to be Euclidean. For the two (real)
scalars this leads to the coset
\begin{equation}
\frac{SL(2,\bb{R})}{SO(1,1)}
\end{equation}
parameterised by
\begin{align}\label{ifactor}
  {\mathcal{M}}_E = e^{{b \phi /2}} \left(
  \begin{array}{cc} - \tfrac{1}{4} b^2 \chi^2+e^{-b \phi} & { \tfrac{1}{2} i
b \chi} \\
{\tfrac{1}{2} i b \chi} & 1 \end{array}
  \right) \,,
\end{align}
where $\phi$ and $\chi$ are both real\footnote{Note the occurrence of factors of $i$ in
(\ref{ifactor}). What matters is that the Lagrangian (\ref{EuclideanAction}) and the
transformation rules of the scalars (\ref{SL2R}) are real.}. The corresponding Euclidean
Lagrangian is
\begin{equation}
\mathcal{L}_E
  = \tfrac{1}{2} \sqrt{g} \,
    [ {R} + b^{-2} \text{Tr}(\partial \mathcal{M}_E \, \partial
\mathcal{M}^{-1}_E) ]
  = \tfrac{1}{2} \sqrt{g} \,
    [ {R} -\tfrac{1}{2} (\partial {\phi})^2
    +\tfrac{1}{2} e^{b {\phi}} (\partial {\chi})^2 ] \, ,
\label{EuclideanAction}
\end{equation}
with all fields real. For $b=2$ and $D=10$ this is the gravity-scalar part of IIB
supergravity after a Wick rotation, i.e.~in Euclidean space. Again, compactifications of
the $D=10$ theory can give rise to other values of $b$. The $SL(2,\bb{R})$ symmetry for
the Euclidean case acts as (see also \cite{Einhorn:2002sj}):
\begin{equation}
  \mathcal{{M}}_E \rightarrow \Omega_E \, \mathcal{{M}}_E \, \Omega_E^T
  ~~~~ \text{with} ~~~
  \Omega_E = \left( \begin{array}{cc} a&ib\\-ic&d \end{array} \right) \,,
\label{SL2R}
\end{equation}
with $a,b,c,d$ real parameters satisfying $ad-bc=1$.

Given the $SL(2,\bb{R})$ symmetry of the field equations, there
are corresponding currents. In the Euclidean case the
$SL(2,\bb{R})$ currents are given by the matrix, see
e.g.~\cite{Meessen:1998qm},
\begin{equation}
J_\mu=(\partial_\mu
\mathcal{M}_E)\mathcal{M}_E^{-1}=\left(\begin{matrix}
j_{\mu}^{(3)}& i\,j_{\mu}^{(+)} \\
i\,j_{\mu}^{(-)}&-j_{\mu}^{(3)}
\end{matrix}\right) \,,
\label{SL2Rcurrents}
\end{equation}
with the following components:
\begin{align}
  j_\mu^{(3)}&=\tfrac{1}{2}\,e^{b\phi}\partial_\mu (e^{-b\phi}- \tfrac{1}{4}
b^2 \chi^2)\,, \qquad
  j_\mu^{(-)}=\tfrac{1}{2}  b e^{b \phi}\partial_\mu \chi\,, \notag \\
  j_\mu^{(+)}&= -b \chi j_\mu^{(3)}+(e^{-b\phi}-\tfrac{1}{4} b^2
\chi^2)j_\mu^{(-)} \,.
\end{align}
We use here a basis where $j_0^{(3)}$ generates a $SO(1,1)$
subgroup of $SL(2,\mathbb{R})$ while $j_0^{(\pm)}$ each generate a
differently embedded $\mathbb{R}$ subgroup of $SL(2,\mathbb{R})$.
The currents \eqref{SL2Rcurrents} satisfy the following field
equations and Bianchi identities:
\begin{align}
  \nabla_\mu J^\mu = 0 \,, \qquad \partial_{[\mu} J_{\nu]} =
J_{[\mu} J_{\nu]} \,.
\end{align}
The first equation corresponds to the field equations of the
Lagrangian \eqref{EuclideanAction} and the second equation follows
from the definition \eqref{SL2Rcurrents}.

Using Stokes' theorem the electric charges of a solution are
obtained by integrating the currents over a $(D-1)$-sphere. We
define our charge matrix as follows:
\begin{equation}
  Q_E=\frac{(2\,(D-1)\,(D-2))^{-1/2}}{b\,\mathcal{V}ol(S^{D-1})}
\int_{S^{D-1}}J_\mu n^\mu\,, \label{charges}
\end{equation}
where $n^\mu$ is an outward directed unit vector. Under an
$SL(2,\mathbb{R})$ transformation \eqref{SL2R} the corresponding
charge matrix transforms as
\begin{align}
  Q_E \rightarrow \Omega_E \, Q_E \, \Omega_E^{-1} \,.
\label{SL2Rtransformations}
\end{align}
Note that the determinant of $Q_E$ is invariant under
$SL(2,\mathbb{R})$. Thus solutions with different values of
$\det(Q_E)$ can never be related via
$SL(2,\bb{R})$-transformations. As discussed in the introduction
the cases $\det(Q_E)=0, \det(Q_E)>0$ and $\det(Q_E)<0$ describe
the three different conjugacy classes of $SL(2,\bb{R})$.

\section{Instanton Solutions} \label{solution}

In this section we will consider instanton-like solutions to the bulk equations of
motion. Issues like boundary terms and values of the action are postponed to section 6.

\subsection{Bulk Solutions}

We consider the Euclidean gravity-dilaton-axion system in $D \geq
3$ dimensions given by the Lagrangian (with arbitrary dilaton
coupling parameter $b$)
\begin{equation}
\mathcal{L}_E
  = \tfrac{1}{2} \sqrt{g} \,
    [ {R} -\tfrac{1}{2}\,(\partial {\phi})^2
    +\tfrac{1}{2}\,e^{b {\phi}} (\partial {\chi})^2 ] \, ,
\label{EuclideanAction2}
\end{equation}
and search for generalised D-instanton solutions with manifest
$SO(D)$ symmetry of the form\footnote{Note that by using
reparameterisations of $r$ one can obtain different, but
equivalent, forms of the metric in which the $SO(D)$ symmetry is
non-manifest, in particular
 \begin{align}
  {ds}^2 & = e^{2 \, B(r)} (e^{-2 \, f(r)} dr^2 + r^2 d\Omega_{D-1}^2) \,,
 \end{align}
in analogy to what we will encounter later, see \eqref{branes}. We
choose to take as our starting point a conformally flat metric,
i.e.~$f(r)=0$.}
\begin{align}
  {ds}^2 & = e^{2\,B(r)} (dr^2 + r^2 d\Omega_{D-1}^2) \,, \qquad
\phi=\phi(r) \,, \qquad \chi=\chi(r)\,.
  \label{instanton}
\end{align}
The standard D-instanton solution \cite{Gibbons:1996vg} is
obtained for the special case that $B(r)$ is constant. In order to
obtain an $SO(D)$ symmetric generalised D-instanton solution, we
allow for a non-constant $B(r)$ and solve the field equations
following from the Euclidean action (\ref{EuclideanAction2}),
which read
\begin{align}
R_{\mu\nu} & = \frac{1}{2}\,\partial_\mu \phi\partial_\nu \phi -
\frac{1}{2}\,e^{b\phi}\partial_\mu \chi\partial_\nu
\chi\,,\notag \\
0&=\partial_\mu \left(\sqrt{g}g^{\mu\nu}e^{b\phi}\partial_\nu
\chi\right)\,,
\notag \\
0&=\frac{b}{2}\,e^{b\phi}(\partial \chi)^2
+\frac{1}{\sqrt{g}}\,\partial_\mu
\left(\sqrt{g}g^{\mu\nu}\partial_\nu
\phi\right)\,\label{FieldEqs}.
\end{align}
The expression for the Ricci tensor for the Ansatz
(\ref{instanton}) is given by
\begin{align}
R_{rr} & = -(D-1)\,\left(B''(r)+\frac{B'(r)}{r}\right)\,, \notag \\
R_{\theta\theta} & =-e^{-2\,B(r)}\,
g_{\theta\theta}\,[B''(r)+(D-2)\,B'(r)^2+(2\,D-3)\,\frac{B'(r)}{r}]\,,
\end{align}
where the prime denotes differentiation with respect to $r$ and $\theta $ denote the
angular coordinates. In addition to the $SL(2,\mathbb{R})$ symmetry these field equations
are invariant under a constant Weyl rescaling of the metric\footnote{In contrast to
$SL(2,\mathbb{R})$, the constant Weyl rescaling symmetry is broken by $O(\alpha^\prime)$
corrections.}
 \begin{align}
  g_{\mu \nu} \rightarrow e^{2 \omega} g_{\mu \nu} \,. \label{Weyl}
 \end{align}
However, this is only a symmetry of the field equations and not of the action. In our
Ansatz \eqref{instanton}, this has the effect of shifting $B$ with a constant, i.e.~$B
\rightarrow B+\omega$.

One can consider the angular component of the Einstein equation of \eqref{FieldEqs} to
solve for $B(r)$. Having solved for $B(r)$ the expressions for the dilaton and axion
scalars can be obtained from the remaining two equations of \eqref{FieldEqs}. We thus
obtain the following solution\footnote{For practical purposes we omit an overall $\pm$
sign corresponding to the $Z_2$ symmetry of the axion, corresponding to the choice
between instanton and anti-instanton. This sign affects some signs in the
$SL(2,\mathbb{R})$ charges of the solution, but does not change its conjugacy class.} for
$B(r), \phi(r)$ and $\chi(r)$, which extends the solution given in \cite{Einhorn:2000ct}
to arbitrary $b$:
\begin{align}
e^{(D-2)\,B(r)} & = f_{+}(r)\,f_{-}(r)\,,\notag\\
e^{b\phi(r)} & =\left(
\frac{q_{-}}{2\,\mathsf{q}}\,[e^{C_{1}}\,(f_{+}(r)/f_{-}(r))^{b\,c/2}-e^{-C_
{1}}\,(f_{+}(r)/f_{-}(r))^{-b\,c/2}]\right)^2\,,\notag\\
\chi(r) &
=\frac{2}{b\,q_{-}}\,[\mathsf{q}\left(\frac{e^{C_{1}}\,(f_{+}(r)/f_{-}(r))^{
b\,c/2}+e^{-C_{1}}\,(f_{+}(r)/f_{-}(r))^{-b\,c/2}}
{e^{C_{1}}\,(f_{+}(r)/f_{-}(r))^{b\,c/2}-e^{-C_{1}}\,(f_{+}(r)/f_{-}(r))^{-b
\,c/2}}\right)-q_{3}]\, . \label{preSolEq}
\end{align}
The solution is given in terms of the two flat-space harmonic
functions
\begin{align}
f_{\pm}(r) & = 1\pm\frac{\mathsf{q}}{r^{D-2}}\,
\end{align}
and the four integration constants $\mathsf{q}, q_3, q_-$ and
$C_1$. The integration constant $\mathsf{q}$ is defined as the
square root of ${\mathsf{q}^2}$, which is an integration constant
that can be positive, zero or negative\footnote{Note that this
implies that the solution (\ref{preSolEq}) is not manifestly real,
since $\mathsf{q}$ can be imaginary.  Below, we discuss this issue
separately for the three cases $\mathsf{q}^2$ positive, negative
or zero.}. Finally, the constant $c$ is given by
\begin{align}
 c & = \sqrt{\frac{2(D-1)}{(D-2)}} \,. \label{ceq}
\end{align}
Note that the metric, specified by $B(r)$ given in
\eqref{preSolEq},
 only depends on the product of $f_+$ and $f_-$ whereas the
scalars only depend on the quotient of $f_+$ and $f_-$. This
reflects the presence of the scale symmetry \eqref{Weyl}, whose
effect is to scale both $f_{\pm}$ with the same factor. The
constants $\mathsf{q}^2$ and $q_-$ occur with inverse powers and
have been taken non-zero in the above solution. Below, we will see
that sending them to zero yields interesting limits.

The solution \eqref{preSolEq} carries electric $SL(2,\mathbb{R})$
charges given by
\begin{equation}
  Q_E= \left(\begin{matrix} q_3 & i q_+ \\ i q_- & - q_3
\end{matrix}\right)\,,
  \label{electric-charge-matrix}
\end{equation}
where we have defined the dependent integration constant $q_+$ via
\begin{align}
  \mathsf{q}^2 = -q_+ q_- + q_3{}^2 = - \det(Q_E) \,.
\end{align}
Thus, the solution (\ref{preSolEq}) has general $SL(2,\mathbb{R})$
charges $(q_+,q_-,q_3)$.

The appearance of the four independent integration constants,
$\mathsf{q}^2$, $q_-$, $q_3$ and $C_1$, can be understood as
follows. As can be inferred from the solution \eqref{preSolEq},
the constant $q_3$ corresponds to the freedom to apply
$\mathbb{R}$ transformations, which shift the axion. Similarly,
the constant $q_-$ corresponds to $SO(1,1)$ transformations, which
scale the axion and shift the dilaton. By applying such
transformations one can shift $q_3$ with arbitrary numbers while
$q_-$ can be rescaled with a positive number. The constant $C_1$
is shifted as follows
\begin{equation}
C_1 \rightarrow C_1 - 2\,\lambda\,\mathsf{q}\label{C1-transf}
\end{equation}
under the $SL(2,\mathbb{R})$ transformation, with parameter
$\lambda$, whose generator is given by the electric charge matrix:
 \begin{align}
  \Omega_E = \exp(\lambda\,Q_E) \,. \label{C1-transformation}
 \end{align}
Since $Q_E$ is invariant under such transformations, see
 \eqref{SL2Rtransformations}, while $C_1$ is
shifted, this explains why $C_1$ does not appear in \eqref{electric-charge-matrix}. The
remaining constant, $\mathsf{q}^2$, is invariant under $SL(2,\mathbb{R})$ and thus does
not correspond to these symmetry transformations. Rather, this constant corresponds to
the freedom to perform rescalings of the metric \eqref{Weyl}. To retain a metric that
asymptotically goes to $1$, this must be combined with an appropriate rescaling of $r$.
The resulting effect of this transformation is a rescaling of $\mathsf{q}^2$ with a
positive number. One therefore always stays in the same conjugacy class under such
transformations.

The solution \eqref{preSolEq} can be written in a more compact
form by using, instead of the two functions $f_+$ and $f_-$ which
are harmonic over $D$-dimensional flat space, a function $H(r)$
which is harmonic over a conformally flat space with the conformal
factor specified by the function $B(r)$ given in \eqref{preSolEq},
i.e.
\begin{align}\label{box1}
  \frac{\partial}{\partial r} \left( r^{D-1} e^{(D-2)\,B(r)} \frac{\partial
  H(r)}{\partial r} \right) = 0 \,.
\end{align}
The general solution to this equation is of the following form:
\begin{align}
  H(r) \propto \log(f_{+}(r)/f_{-}(r))\, . \label{harmonic}
\end{align}
We can, therefore, rewrite the solutions \eqref{preSolEq} as
follows:
\begin{equation}\label{SolEq}
\boxed{
\begin{aligned}
  ds^2 & = \left(1-\frac{\mathsf{q}^2}{r^{2\,(D-2)}}\right)^{2/(D-2)}\,(dr^2
+ r^2 d
  \Omega_{D-1}^2) \,, \\
  e^{b\,\phi(r)} & =
\left(\frac{q_{-}}{\mathsf{q}}\,\sinh(H(r)+C_1)\right)^2\,,\\
  \chi(r) & =
  \frac{2}{b\,q_{-}}\,(\mathsf{q}\,\coth(H(r)+C_1)-q_3)\,,
\end{aligned}
}
\end{equation}
where
\begin{align}
  H(r)= \frac{b\,c}{2}\,\log(f_{+}(r)/f_{-}(r))\,.
\end{align}

The solutions \eqref{SolEq} are valid both for $\mathsf{q}^2$
positive, negative and zero. Below we discuss the reality and
validity of the solutions for each of these three cases. Note that
we use the Einstein frame.
\begin{itemize}
\item
 $\bf \mathsf{q}^2 >0$:

In this case $\mathsf{q}$ is real and the solution is given by
\eqref{SolEq} with all constants real.
However, the metric poses a problem: it becomes imaginary for
 \begin{equation} \label{rcritical}
  r^{D-2} < r_c^{D-2} = \mathsf{q} \, .
 \end{equation}
One can check that there is a curvature singularity at $r=r_c$.
However, this curvature singularity happens at strong string
coupling:
 \begin{equation}
  e^{\phi(r)} \rightarrow \infty \,, \hskip 2truecm r \rightarrow r_c \, .
 \end{equation}
Between $r=r_c$ and $r=\infty$, $H$ varies between $\infty$ and $0$, and with
an appropriate choice\footnote{According to \eqref{C1-transf}, the
constant $C_1$ can be changed by an $SL(2,\mathbb{R})$ transformation, leading
to singular scalars (but non-singular currents, which are independent
of $C_1$). However, since these are related to regular scalars by
a global $SL(2,\mathbb{R})$ transformation, this does not pose a
problem.} of $C_1$, i.e. a positive value of $C_1$, the scalars have no
further singularities in this domain.
Thus one might hope to have a modification of this solution by higher-order contributions
to the effective action of IIB string theory \cite{Einhorn:2002am}. Alternatively, one
can consider the possible resolution of this singularity upon uplifting. In the next
section, we will see that this indeed happens for the special case of
\begin{equation}
b = \sqrt{\frac{2(D-2)}{D-1}} \,,
\end{equation}
equivalent to $bc = 2$.

In the case with $\mathsf{q}^2 > 0$, there is an interesting limit
in which $q_- \rightarrow 0$. For generical values of the other
three constants, this yields a non-sensible solution with infinite
scalars. To avoid this, one must simultaneously impose
 \begin{align} \label{q-minus-vanishing}
  C_1 \rightarrow - \log(\frac{q_-}{2 \mathsf{q}}) \,, \qquad
  q_3 \rightarrow \mathsf{q} - \frac{q_+ q_-}{2 \mathsf{q}} \,, \qquad
  q_- \rightarrow 0 \,.
 \end{align}
This yields a well-defined limit, in which the scalars read
 \begin{align}\label{Schwarzschild}
  e^{\phi/c} = \frac{f_+}{f_-} \,, \qquad
  \chi = \frac{-q_+}{b \mathsf{q}} \,,
 \end{align}
while the metric is unaffected and given by \eqref{preSolEq}. Note
that in this limit the dilaton becomes independent of $b$: when
the axion is constant, the dilaton coupling drops out of the field
equations. In this limit, one is left with two independent
integration constants, $q_+$ and $\mathsf{q}^2$. The range of
validity of this solution is equal to that of the above solution
with $q_- \neq 0$: it is well-defined for $r>r_c$, while at $r =
r_c$ the metric has a singularity and the dilaton blows up. We
will find that this singularity is resolved upon uplifting for all
values of $bc \geq 2$.

\item $\bf \mathsf{q}^2 =0$

 We now consider the limit $\mathsf{q}^2 \rightarrow 0$ of
the general solution \eqref{SolEq}. Taking this limit for generic
values of $C_1$, one sees that $e^{\phi(r)} \rightarrow \infty$
for all $r$. The only way to avoid this bad behaviour is to have
$C_1 \rightarrow 0$, as $\mathsf{q}^2 \rightarrow 0$. Thus, to
obtain a well-defined limit, we simultaneously take
\begin{align}\label{limit}
  C_1 \rightarrow g_s^{b/2}\,\frac{\mathsf{q}}{q_{-}} \,, \qquad
  \mathsf{q}^2 \rightarrow 0 \,.
\end{align}
The constant $g_s$ is assumed positive and will correspond to the
value of $e^{\phi(r)}$ at $r=\infty$. Taking the limit
\eqref{limit} of the general solution \eqref{SolEq} yields the
extremal solution:
\begin{equation}
\boxed{
\begin{aligned}
  ds^2 = dr^2+r^2\,d\Omega_{D-1}^2 \,, \qquad
  e^{b\,\phi(r)/2} = h \qquad \chi(r) = \frac{2}{b}\,(h^{-1} -
\frac{q_3}{q_-}) \,,   \label{instlimEq}
\end{aligned}
}
\end{equation}
where $h(r)$ is the harmonic function:
\begin{align}
  h(r) = g_s^{b/2} +  \frac{b\,c\,q_-}{r^{D-2}} \, ,
\end{align}
This is the extremal D-instanton solution of \cite{Gibbons:1996vg}. This solution is
regular over the range $0 < r < \infty$ provided one takes both $g_s$ and $b\,c\,q_-$
positive; at $r=0$ however, the harmonic function blows up and the scalars are singular.

\item $\bf \mathsf{q}^2 <0$:

In this case $\mathsf{q}$ is imaginary. To obtain a real solution
we must take $C_1$ to be imaginary. We therefore redefine
\begin{equation}
\mathsf{q}\rightarrow i\,\mathsf{\tilde{q}} \hskip 2truecm C_1
\rightarrow i\,\tilde{C_1}\, ,
\end{equation}
such that $\mathsf{\tilde{q}}$ and $\tilde{C_1}$ are real. One can
now rewrite the solution \eqref{SolEq} by using the
relation\footnote{Here we have used the general relation
$\log((1+x)/(1-x)) = 2\,{\rm arctanh}(x)$.}
 \begin{equation}
  \log(f_{+}/f_{-}) = 2 \,{\rm arctanh}(\mathsf{q}/r^{D-2})\, ,
 \end{equation}
and, next, replacing the hyperbolic trigonometric functions by
trigonometric ones in such a way that no imaginary quantities
appear. We thus find that, for $\mathsf{q}^2 <0$, the general
solution \eqref{SolEq} takes the following form:
\begin{equation}
\boxed{
\begin{aligned}
  ds^2 & =
(1+\frac{\mathsf{\tilde{q}}^2}{r^{2\,(D-2)}})^{2/(D-2)}\,(dr^2+r^2\,d\Omega_
{D-1}^2)\,,\\
  e^{b \phi(r)} & = \left(\frac{q_{-}}{\mathsf{\tilde{q}}}\,
  \sin(b\,c\,\arctan(\frac{\mathsf{\tilde{q}
}}{r^{D-2}})+\tilde{C_1}) \right)^2 \,,\\
  \chi(r) & =\frac{2}{b\,q_{-}}\,(\mathsf{\tilde{q}}\,
  \cot(b\,c\,\arctan(\frac{\mathsf{\tilde{q}}}
{r^{D-2}})+\tilde{C_1})-q_3)\,.
 \label{qminussol}
\end{aligned}
}
\end{equation}
The metric and curvature are well behaved over the range $0<r<\infty$. However, the
scalars can only be non-singular over the same range by an appropriate choice of
$\tilde{C}_1$ provided that $bc < 2$. This can be seen as follows. The $\arctan$ varies
over a range of $\pi/2$ when $r$ goes from $0$ to $\infty$. It is multiplied by $bc$ and
thus the argument of the sine varies over a range of more than $\pi$ if $bc > 2$.
Therefore, for $bc>2$ there is always a point $r_c$ such that $\chi \rightarrow \infty$
as $r \rightarrow r_c$. Note that the breakdown of the solution occurs at weak string
coupling: $e^{\phi} \rightarrow 0$ as $r \rightarrow r_c$. In the next section we will
find that this singularity is not resolved upon uplifting and will correspond to a naked
singularity. The same holds for the liming case of $bc=2$. Therefore the case
$\mathsf{q}^2 <0$ only yields regular instanton solutions for $bc < 2$, together with the
condition that $C_1$ and $C_1+bc\pi/2$ are on the same branch of the cotangent.

\end{itemize}

\subsection{Wormhole Geometries} \label{wgeometry}

\begin{figure}[ht]
\centerline{\epsfig{file=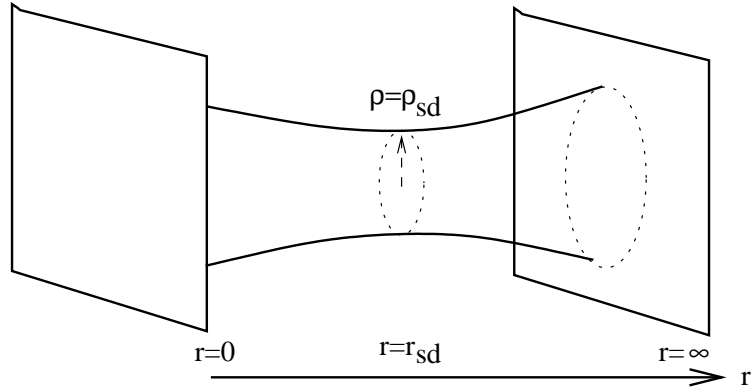,width=.6\textwidth}}
 \caption{\it The geometry of a wormhole. The two asymptotically flat
regions at $r=0$ and $r=\infty$ are
 connected via a neck with a minimal physical radius $\rho_{\text{sd}}$ at
the self-dual radius $r_{\text{sd}}$.}
 \label{fig:wormhole}
\end{figure}

It is  known \cite{Gibbons:1996vg} that the standard D-instanton,
i.e.~$D=10, b=2$, in string frame has the geometry of a wormhole,
i.e. it has two asymptotically flat regions connected by a neck,
see figure \ref{fig:wormhole}. It will therefore be interesting to
investigate whether there exists frames in which the non-extremal
instantons also have the geometries of wormholes.

We consider a general wormhole metric of the form
\begin{align}\label{genwormholemetric}
  ds^2 & = f(r)^{2/(D-2)} \, (dr^2 + r^2 d\Omega^2) \,, \qquad
  f(r) = \alpha+\beta r^{2-D}+\gamma r^{4-2D} \,,
\end{align}
where $\alpha$, $\beta$ and $\gamma$ are constants. The metric has
a $\mathbb{Z}_2$ isometry corresponding to the transformation
$r^{D-2} \rightarrow \gamma \, r^{2-D} / \alpha$ which
interchanges the two asymptotically flat regions. The physical
radius $\rho$ is the square root of the coefficient of the angular
part of the metric, given by $\rho^{D-2} = f(r) r^{D-2}$. The
minimum of this physical radius of the neck occurs at the fixed
point of the transformation above, i.e. at the so-called self-dual
radius $r_{\text{sd}}^{D-2} = \sqrt{\gamma/\alpha}$, and is given
by $\rho_{\text{sd}}^{D-2} = 2\sqrt{\alpha\gamma}+\beta$. We will
now study the three conjugacy classes in order to see for each
case if there exists a frame\footnote{In arbitrary dimension one
can define three different frames as follows: in the Einstein
frame the Einstein-Hilbert term has no dilaton factor, in the
string frame the kinetic term for the axionic field strength comes
without a dilaton factor (like all Ramond-Ramond field strengths)
and in the dual frame the Einstein-Hilbert term, the dilaton
kinetic term and the kinetic term for the dual field strength
(i.e. $F^2_{D-p-2}$ for the frame dual to a $p$-brane) come with
the same dilaton factor (see e.g.
\cite{Boonstra:1998mp,Behrndt:1999mk} for a more detailed
discussion).} in which the metric takes the form
\eqref{genwormholemetric}.

\begin{itemize}
\item $\bf \mathsf{q}^2 >0$: As we will see in section \ref{wormhole}, the
appropriate frame in this case is the frame dual to the instanton,
i.e. the $(D-3)$-brane frame, given by
 \begin{align}
  g_{\mu \nu}^{\text{dual}} = e^{b \phi/(D-2)} \, g_{\mu \nu}^{\text{E}} \,.
  \label{dual-frame}
 \end{align}
In the special case of $b \, c = 2$, the metric takes the form
\eqref{genwormholemetric} in the dual frame with
\begin{equation}
  f(r) =  \frac{q_-}{\mathsf{q}} \sinh(C_1) + 2 q_- \cosh(C_1) r^{2-D} + q_-
\mathsf{q} \sinh(C_1) r^{4-2D} \,.
\end{equation}
This gives the self-dual radius $r_{\rm sd}$ and the minimal
physical radius $\rho_{\rm sd}$
\begin{align}
 r_{\text{sd}}^{D-2}=\mathsf{q} \,, \qquad
 \rho_{\text{sd}}^{D-2} = 2 q_- e^{C_1} \,.
 \end{align}
Note that the self-dual radius $r_{\text{sd}}$ coincides with the critical radius $r_c$
of the previous section: the curvature singularity in Einstein frame becomes the center
of the wormhole in the dual frame. The limit $q_-\rightarrow 0$, with appropriate scaling
of $C_1$ as given in \eqref{q-minus-vanishing}, yields $\rho_{\text{sd}}^{D-2} =
4\mathsf{q}$. For generic values of $b c$, the instanton metrics can not be written in
the form \eqref{genwormholemetric} in any frame.

\item $\bf \mathsf{q}^2 =0$: It turns out that for any value of $b$ the
wormhole geometry is made manifest by going to the string frame
 \begin{align}
  g_{\mu \nu}^{\text{str}} = e^{2 b \phi/(D-2)} \, g_{\mu \nu}^{\text{E}}
\,.
 \label{string-frame}
 \end{align}
In this frame, the metric is given by \eqref{genwormholemetric}
with
\begin{equation}
 f(r) =  g_s^{b}+2 b c q_- g_s^{b/2}r^{2-D}+(b c q_-)^2 r^{4-2D} \,.
\end{equation}
This gives the self-dual and minimal physical radii
 \begin{align}
 r_{\text{sd}}^{D-2}= b c q_-/g_s^{b/2} \,, \qquad
 \rho_{\text{sd}}^{D-2} = 4 b c q_- g_s^{b/2} \,.
 \end{align}

\item $\bf \mathsf{q}^2 <0$: Here the metric has the appropriate form
already in Einstein frame so from \eqref{qminussol} we get, for
any value of $b$,
 \begin{align}
 r_{\text{sd}}^{D-2}=\mathsf{\tilde{q}} \,, \qquad
 \rho_{\text{sd}}^{D-2} = 2\mathsf{\tilde{q}} \,.
 \end{align}

\end{itemize}
We thus see that for all three conjugacy classes there exists
frames in which the solutions have the geometries of wormholes.

\subsection{Instanton Solutions with Multiple Dilatons}

We will now consider extensions of the instanton solution of the previous sections, which
is carried by the $SL(2,\mathbb{R})$ scalars $\phi$ and $\chi$. We will extend this
system with $n$ dilatons $\varphi_\alpha$\ $(\alpha = 1, \ldots ,n)$, which are
$SL(2,\mathbb{R})$ singlets and do not couple to the axion (this can always be achieved
by field redefinitions provided one allows for an arbitrary dilaton coupling $b$ to the
original dilaton $\phi$). We will call the corresponding solution a multi-dilaton
instanton. The multi-dilaton action is given by
\begin{equation}
\mathcal{L}_E
  = \tfrac{1}{2} \sqrt{g} \,
    [ {R} - \tfrac{1}{2}\,\sum_{\alpha=1}^n (\partial\varphi_\alpha)^2
    - \tfrac{1}{2}\,(\partial {\phi})^2
    +\tfrac{1}{2}\,e^{b {\phi}}\,(\partial {\chi})^2 ] \, .
\label{genaction}
\end{equation}
with field equations \eqref{FieldEqs} plus $n$ equations,
requiring $\varphi_\alpha$ to be harmonic in the curved space. The
case of one extra dilaton was considered in \cite{Cremmer:1998em}.

The solution to this system has the same metric as given in \eqref{preSolEq}, see
also~\cite{Cremmer:1998em}. Then the extra dilatons $\varphi_\alpha$ satisfy a
d'Alembertian equation in a conformally flat background specified by $B(r)$ as given in
\eqref{preSolEq}:
\begin{align}\label{box2}
  \frac{\partial}{\partial r} \left( r^{D-1} e^{(D-2)\,B(r)} \frac{\partial
  \varphi(r)}{\partial r} \right) = 0 \,.
\end{align}
This equation is solved by the harmonic function as given in
\eqref{harmonic}, yielding dilatons given by
\begin{equation}\label{solutionbox}
  \varphi_\alpha = \nu_\alpha + \mu_\alpha \log \left(\frac{f_+(r)}{f_-(r)}\right) \, ,
\end{equation}
with $2n$ integrations constants $\nu_\alpha$ and $\mu_\alpha$.

Of course, due to the presence of the extra dilatons $\varphi_\alpha$, the Einstein
equation in \eqref{FieldEqs} is modified. It turns out that the contribution of
$\varphi_\alpha$ to the energy-momentum tensor is cancelled by similar
$\mu_\alpha$-dependent contributions of the dilaton $\phi$ and the axion $\chi$ to the
energy-momentum tensor. Since all $\mu_\alpha$-dependent contributions of the dilatons
and the axion to the energy-momentum tensor cancel against each other, this extension
allows for a $\mu_\alpha$-independent metric.

\section{Uplift to Black Holes}\label{wormhole}

\subsection{Kaluza-Klein Reduction} \label{reduction}

In this section we consider the possible higher-dimensional origin
of the Euclidean system \eqref{EuclideanAction2} as a consistent
truncation of the $(D+1)$-dimensional Lagrangian, defined over
Minkowski space,
 \begin{align}
  \mathcal{L}_{D+1} = \sqrt{-\hat{g}}\,[\hat{R} - \tfrac{1}{2}\,(\partial
\hat{\phi})^2
   - \tfrac{1}{4}\,e^{a \hat{\phi}}\,\hat{F}^2 ] \,,
  \label{Einstein-Maxwell-dilaton}
 \end{align}
with the rank-2 field strength $\hat{F}=d \hat{A}$. It consists of
an Einstein-Hilbert term (for a metric of Lorentzian signature), a
dilaton kinetic term and a kinetic term for a vector potential
with arbitrary dilaton coupling, parameterised by $a$. The
corresponding $\Delta$ value \cite{Lu:1995cs} is given by
 \begin{align}
  \Delta = a^2 + \frac{2\,(D-2)}{D-1} \,,
  \label{Delta}
 \end{align}
which characterises the dilaton coupling in $D+1$ dimensions.

The reduction Ansatz over the time coordinate is
 \begin{align}
  \hat{ds}^2 = e^{2 \alpha \varphi}\,ds^2 - e^{2 \beta \varphi}\,dt^2 \,,
\qquad \hat{A} =
  \chi\,dt \,, \qquad \hat{\phi} = \phi \,,
 \label{reduction-Ansatz}
 \end{align}
with the constants
 \begin{align}
  \alpha^2 = \frac{1}{2\,(D-1)\,(D-2)} \,, \qquad \beta = - (D-2)\,\alpha
\,,
 \label{constants-circle}
 \end{align}
which are chosen such as to obtain the Einstein frame in the lower dimension with
appropriate normalisation of the dilaton $\varphi$. Note that the dilaton factor in front
of the spatial part of the metric $\hat{g}_{\mu \nu}$ coincides, for $bc=2$, with the
dual frame defined in section \ref{wgeometry}.

With the Ansatz as above, the Einstein-Maxwell-dilaton system
reduces to the $D$-dimensional Euclidean system
 \begin{align}
  \mathcal{L}_D = \sqrt{-g}\,[ R -\tfrac{1}{2}\,(\partial \phi)^2
  -\tfrac{1}{2}\,(\partial \varphi)^2 + \tfrac{1}{2}\,e^{a\,\phi -
2\,\beta\,\varphi}\,(\partial\chi)^2]
  \,.
 \label{action-2dilatons}
 \end{align}
Next, we perform a field redefinition corresponding to a rotation
in the $(\phi, \varphi)$-plane such that we obtain
 \begin{align}
  \mathcal{L}_D = \sqrt{-g}\,[ R
-\tfrac{1}{2}\,(\partial\tilde{\phi})^2
  -\tfrac{1}{2}\,(\partial\tilde{\varphi})^2 +
\tfrac{1}{2}\,e^{b\,\tilde{\phi}}\,(\partial\chi)^2]
  \,,
 \label{action-2dilatons-a}
 \end{align}
with dilaton coupling $b$ given by
 \begin{align}
  b = \sqrt{a^2 + \frac{2(D-2)}{D-1}} \,.
 \end{align}
The corresponding value of $\Delta$ is equal to the original value
\eqref{Delta}. This system can be truncated to the one we are
considering by setting $\tilde{\varphi} = 0$.

Therefore, the system that we consider in section \ref{solution} has a higher-dimensional
origin if the dilaton coupling satisfies $bc \geq 2$ or
 \begin{align}
  b \geq \sqrt{\frac{2(D-2)}{D-1}} \,.
 \end{align}
The case which saturates the inequality, i.e.~$a=0$, can be
uplifted to an Einstein-Maxwell system without the dilaton
${\hat\phi}$. For $bc > 2$ one needs to include an explicit
dilaton $\hat\phi$ in the higher-dimensional system, i.e.~one must
consider the Einstein-Maxwell-dilaton system
\eqref{Einstein-Maxwell-dilaton} with $a \neq 0$. Note that in
string theory toroidal reductions, under which the combination
$\Delta$ is preserved, only lead to values of $b$ with $bc \ge 2$.

Since the Euclidean gravity-axion-dilaton system we are considering can be obtained as a
consistent truncation of the higher-dimensional Minkowskian Einstein-Maxwell-dilaton
system \eqref{Einstein-Maxwell-dilaton}, it is natural to look for a higher-dimensional
origin of the non-extremal instanton solutions within this system. In the following two
sections we consider the cases $bc=2$ and $bc>2$ separately. The instantons with $bc<2$
have no higher-dimensional origin from toroidal reduction.

\subsection{Reissner-Nordstr\"om Black Holes: $bc = 2$}

It is not difficult to see that for $bc=2$ the generalised
instanton solutions uplift to the $(D+1)$-dimensional
Reissner-Nordstr\"om (RN) black hole solution
\begin{align}
ds^2 & = -g_+(\rho)\,g_-(\rho)\,dt^2+\frac{d\rho^2}{g_+(\rho)\,g_-(\rho)}+ \rho^2
d\Omega^2_{D-1}\,, \quad F_{t\rho}=-\partial_{\rho}A_{t}=(D-2)\,c\,\frac{Q}{\rho^{D-1}}\,
,
\end{align}
where
\begin{align}\label{outer}
g_{\pm}(\rho) &=1-\frac{\rho_{\pm}^{D-2}}{\rho^{D-2}}\,, \qquad
\rho_{\pm}^{D-2}=M\pm\sqrt{M^2-Q^2}\,,
\end{align}
and $Q$ ($M$) is the charge (mass) of the black hole. The RN black
hole has naked singularities for $M^2 < Q^2$, while these are
cloaked for $M^2 \ge Q^2$, yielding a physically acceptable
space-time. Note that the coordinate $\rho$ coincides with the
physical radius of the previous section, for which the angular
part of the metric $d\Omega^2_{D-1}$ is multiplied by $\rho^2$.

In order to establish the precise relation between the charge $Q$
and the mass $M$ of the RN black hole and the $SL(2,\mathbb{R})$
charges of the $bc=2$ instanton solutions given in \eqref{SolEq}
we must first cast the RN  metric in isotropic form as follows:
\begin{align}
ds^2&=-\frac{g(r)}{\rho(r)^{2\,(D-2)}}\,dt^2+\frac{\rho(r)^2}{r^2}\,(dr^2+r^
2\,d\Omega^2_{D-1})\,, \label{RN}
\end{align}
where
\begin{align}
\rho(r)&=(r^{D-2}+M+\frac{M^2-Q^2}{4\,r^{D-2}})^{1/(D-2)}\,,
\qquad g(r)=(r^{D-2}-\frac{M^2-Q^2}{4\,r^{D-2}})^2\,.
\end{align}
To relate the instanton and black hole solutions we need to choose
proper boundary conditions for the instanton solutions
\eqref{SolEq}, which are implied by the boundary conditions of the
RN black hole:
\begin{align}
 \begin{array}{l}
  \lim_{r\rightarrow\infty} g_{tt} = -1 \,, \\ \lim_{r\rightarrow\infty}
A_{t} = 0 \,,
 \end{array} \qquad
 \Longleftrightarrow \qquad
 \begin{array}{c}
  \lim_{r\rightarrow\infty}e^\phi = 1 \,, \\ \lim_{r\rightarrow\infty}\chi =
0 \,.
 \end{array}
\end{align}
This fixes the constants $C_1$ and one of the three
$SL(2,\mathbb{R})$ charges $q_3$ in \eqref{SolEq} as follows:
\begin{align}
C_1&={\rm arcsinh}(\frac{\mathsf{q}}{q_{-}}) \,, \qquad
q_3=\mathsf{q}\,\coth(C_1)=\sqrt{\mathsf{q}^2+q_{-}^2}\,.
\end{align}

The relation between the charge $Q$ and the mass $M$ of the RN
black hole and the two unfixed $SL(2,\mathbb{R})$ charges $q_-$
and $\mathsf{q}^2$ is:
\begin{align}
Q&= -2\,q_{-} \,, \qquad M=2\,\sqrt{\mathsf{q}^2+q_{-}^2}\,,
\label{RN-instanton-relation}
\end{align}
such that
\begin{equation}\label{q2}
\mathsf{q}^2 = \frac{M^2-Q^2}{4}\, .
\end{equation}

{}From \eqref{q2} we see that the physically acceptable non-extremal RN black holes with
$M^2 \ge Q^2$ coincide with the uplifted instanton solutions in the $\mathsf{q}^2=0$ and
$\mathsf{q}^2>0$ conjugacy classes:
\begin{align}
M^2 > Q^2 \qquad &\Leftrightarrow \qquad \mathsf{q}^2 > 0 \,, \notag \\
M^2 = Q^2 \qquad &\Leftrightarrow \qquad \mathsf{q}^2 = 0 \,.
\end{align}
More specifically, we find that the non-extremal (extremal) RN metric in isotropic
coordinates \eqref{RN} reduces to the $\mathsf{q}^2>0$ ($\mathsf{q}^2=0$) instanton solution in the dual
frame metric \eqref{dual-frame}. Note that the $\mathsf{q}^2>0$ instanton has a wormhole geometry
in the dual frame metric. It turns out that the minimal physical radius $\rho_{\rm sd}$
for this case is given by $\rho_{\rm sd} = \rho_+$, where $\rho_+$ is the position of the
outer event horizon given in \eqref{outer}.

\subsection{Interpretation of Instantons as BH
Wormholes}\label{interpretation}

In the previous section we have seen that the non-extremal D-instanton solutions
\eqref{SolEq} in the dual frame metric \eqref{dual-frame} with $b \, c = 2$ and $M^2 \ge
Q^2$ can be viewed as a $t=constant$ space-like section of the RN black hole metric
\eqref{RN}. In the Kruskal-Szeres-like extension of the RN black hole, the spatial part
of the metric \eqref{RN} has the geometry of an Einstein-Rosen bridge or wormhole, which
connects two asymptotically flat regions of space (see \cite{Townsend:1997ku} for a
general introduction to black holes). Indeed, the spatial part of \eqref{RN} has, for
$M^2 > Q^2$, the $\mathbb{Z}_2$ isometry
\begin{align}
  r^{D-2} \rightarrow \frac{M^2-Q^2}{4\,r^{D-2}}\,, \label{Z2-isometry}
\end{align}
which relates each point on one side of the Einstein-Rosen bridge
to a point on the other side.

\begin{figure}[ht]
\centerline{\epsfig{file=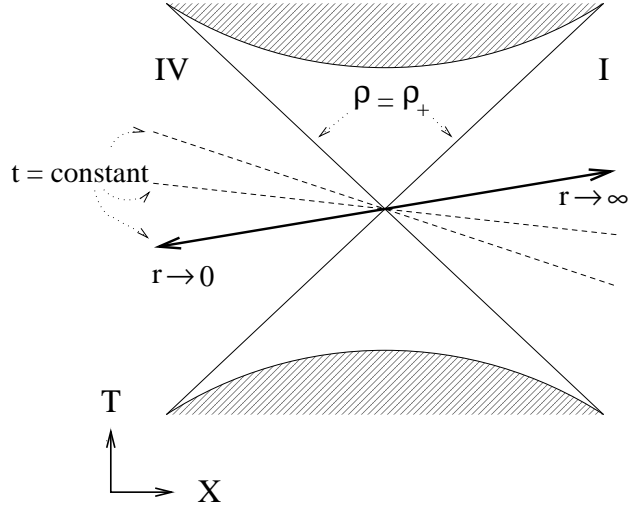,width=.5\textwidth}}
 \caption{\it Schwarzschild black hole in Kruskal-Szeres coordinates.
 Spatial sections with $t=constant$ are space-like lines through the
 origin, going from region $IV$ to region $I$. $T$ and $X$ are the
Kruskal-Szeres time-like and space-like
 directions respectively. The horizons are at $\rho = \rho_+$,
which coincides with the minimal physical radius at the center $\rho = \rho_{\rm sd}$.}
\label{fig:Schwarzschild-wormhole}
\end{figure}

It is instructive to consider the special case of the
Schwarzschild black hole, (i.e.~$Q=0$). Due to
\eqref{RN-instanton-relation}, this corresponds to the uplift of
instantons with $q_- = 0$, i.e.~the solutions given in
\eqref{q-minus-vanishing}. As shown in figure
\ref{fig:Schwarzschild-wormhole}, in the Kruskal-Szeres extension
of the Schwarzschild black hole, every $t=constant$ section of
space time corresponds to a straight space-like line going through
the origin of this coordinate system, with slope determined by the
constant value of $t$.

Notice that on each line, the coordinate $r$ from \eqref{RN} runs
from $r=0$ at the spatial infinity on the left-hand-side, to
$r=\infty$ on the right-hand-side. The fixed point of the
$\mathbb{Z}_2$-isometry \eqref{Z2-isometry} (now with $Q=0$) is
positioned at the center of
figure~\ref{fig:Schwarzschild-wormhole}. The value of $r$ at this
fixed point and the corresponding minimal physical radius is given
by
\begin{equation}
r_{\rm sd}^{D-2} = \tfrac{1}{2} M \, ,\hskip 2truecm \rho_{sd}^{D-2} = 2M \, .
\end{equation}
Note that this value of the physical radius corresponds to the
horizon of the black hole, as can also be seen from
figure~\ref{fig:Schwarzschild-wormhole}. One can make the wormhole
geometry visible by associating to every value of $r$ a
$(D-1)-$sphere. Representing every $(D-1)-$sphere by a circle one
obtains the wormhole picture of figure~\ref{fig:wormhole}.

\begin{figure}[ht]
\centerline{\epsfig{file=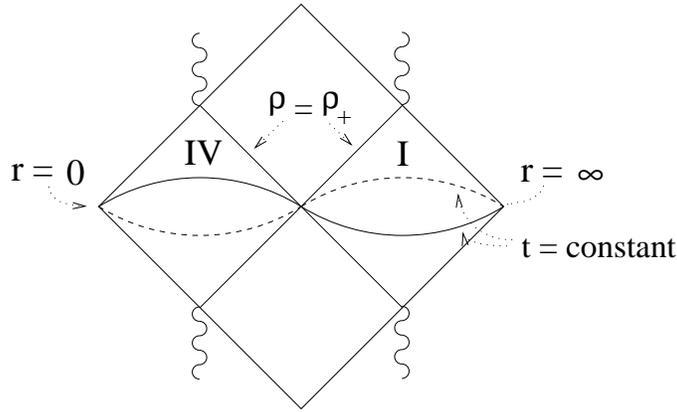,width=.55\textwidth}}
 \caption{\it Carter-Penrose diagram of RN black hole. The lines with $\rho
= \rho_+$ are the horizons, which
 coincide with the minimal physical radius $\rho = \rho_{\text{sd}}$ in the
center.}\label{fig:Penrose}
\end{figure}

In the more general case (i.e. $Q\neq0$), the $t=constant$
sections are still paths connecting two regions of the RN black
hole. To see what these regions correspond to, it is helpful to
draw a Carter-Penrose diagram, see figure~\ref{fig:Penrose}. The
wormhole geometry is qualitatively the same as in the
Schwarzschild case. The position  of the wormhole throat and the
value of the minimal physical radius are given by
\begin{equation}
r_{\rm sd}^{D-2} = \tfrac{1}{4} (M^2-Q^2) \, , \hskip 2truecm \rho_{sd}^{D-2} = M +
\sqrt{M^2-Q^2} \, ,
\end{equation}
which again coincide with the horizon at $\rho = \rho_+$. The curvature singularity of
the D-instanton solutions with $\mathsf{q}^2>0$ \eqref{SolEq} at $r_{\rm c} =
(\mathsf{q})^{1/D-2}$ are resolved in this uplifting and can now be understood as the
usual coordinate singularity of the RN black hole outer event horizons (i.e.
$\rho=\rho_+$, or $r^{2\,(D-2)}=(M^2-Q^2)/4$).

\begin{figure}[ht]
\centerline{\epsfig{file=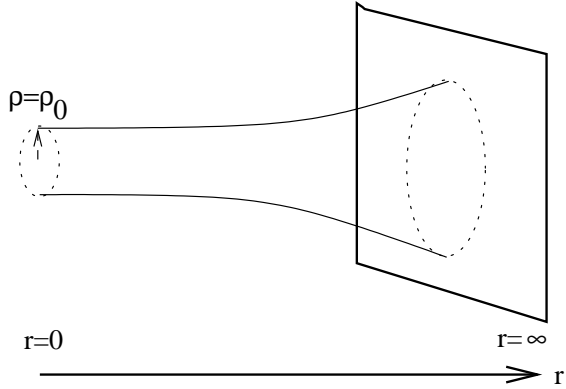,width=.45\textwidth}}
 \caption{\it The geometry of the extremal black hole as a "one-sided"
wormhole
 with minimal physical radius $\rho_{0}$.}\label{fig:one-sided-wormhole}
\end{figure}

The extremal RN black hole (i.e. $M^2=Q^2$) is qualitatively different from the other
cases. As one can see from \eqref{Z2-isometry}, the $\mathbb{Z}_2$-isometry is gone. By
taking the limit $M^2\rightarrow Q^2$ of a non-extremal black hole we see that the
wormhole stretches to an infinitely long throat. The fixed point of the isometry goes to
spatial infinity at $r=0$. This means that the extremal black hole has a "one-sided"
wormhole with a minimal physical radius $\rho_{0}^{D-2}=M$, and the full Kruskal-like
extension is geodesically complete without need for a region $IV$. This situation is
illustrated in figure~\ref{fig:one-sided-wormhole}.

\subsection{Dilatonic Black Holes: $bc>2$}

The instantons with $bc > 2$ uplift to non-extremal {\it dilatonic} black holes,
i.e.~black hole solutions carried by a metric, a vector and a dilaton. In fact, the
uplift is identical to a version of the black hole solution presented in
\cite{Lu:1996hh}. To be more precise, the non-extremal dilatonic black hole solutions of
\cite{Lu:1996hh} contain an extra parameter $\mu$. For generic values of this parameter
the black hole solution is singular\footnote{These (singular) solutions are a
generalisation of the (regular) black holes of \cite{Gibbons:1988ps}.}. One only obtains
a regular solution if\footnote{The parameter $\mathsf{q^2}$ can be identified with the
parameter $k$ of \cite{Lu:1996hh}.} $\mu \sim \mathsf{q}$.

The uplift of the $bc > 2$ instantons equals the $\mu \rightarrow 0$ limit of the
non-extremal black hole solutions of \cite{Lu:1996hh}. Therefore, in contradistinction to
the $bc=2$ case, we obtain a singular black hole solution. This singularity can only be
avoided in two limiting cases. The singularity disappears both in the extremal limit
\eqref{limit} when $\mathsf{q}^2 \rightarrow 0$ and in the Schwarzschild limit
\eqref{Schwarzschild} when $q_-\rightarrow 0$, where the dilaton decouples.

\section{Uplift to $p$-Branes} \label{p-branes}

In section 4 we have discussed the uplift of the instantons of
section 3 to higher-dimensional black hole solutions. It is
therefore natural to consider the uplift to higher-dimensional
$p$-branes. To this end it will be useful to first introduce the
following nomenclature.

Non-extremal deformations of general $p$-branes have been
considered in \cite{Horowitz:1991cd, Lu:1996hh}. These are
solutions of the $(D+p+1)$-dimensional Lagrangian, defined over
Minkowski space,
 \begin{align}
  \mathcal{L}_{D+p+1} = \sqrt{-\hat{g}}\,[\hat{R} - \frac{1}{2}\,(\partial
\hat{\phi})^2
   - \frac{1}{2 \, (p+1)!} \, e^{a \hat{\phi}}\,\hat{G}_{(p+2)}^2 ] \,,
 \end{align}
with the rank-$(p+2)$ field strength $\hat{G}_{(p+2)} = d
\hat{C}_{(p+1)}$. For a $p$-brane in $D+p+1$ dimensions the metric
(in Einstein frame) is of the form
 \begin{align}
  ds^2 = e^{2A} (- e^{2f} dt^2 + dx_p{}^2) + e^{2B} (e^{-2f} dr^2 + r^2
   d \Omega_{D-1}{}^2) \,,
 \label{branes}
 \end{align}
where $A$, $B$ and $f$ are functions that depend on the radial
coordinate $r$ only. It is convenient to introduce the quantity
 \begin{equation}
  X = (p+1) A + (D-3) B\, .
 \end{equation}
The extremal $p$-brane solutions with equal mass and charge, preserving half of the
supersymmetry, are obtained by taking $X=f=0$.

Assuming that $D\ge 3$ there exist two types of non-extremal $p$-brane solutions in the
literature. Following \cite{Lu:1996hh}, we will call them type 1 and type 2 non-extremal
$p$-branes:
 \begin{itemize}
 \item {\bf Type 1 non-extremal $p$-branes:}\ $X = 0$ and $f \neq 0$.

These are the non-extremal black branes of
  \cite{Horowitz:1991cd, Duff:1994ye}. The deformation function $f$ is given by
  \begin{align}
   e^{2f} = 1 - \frac{k}{r^{D-2}} \,,
  \end{align}
where $k$ is the deformation parameter. In a different coordinate frame, with radial
coordinates $\rho$,  these branes can be expressed in terms of the two harmonic functions
  \begin{align}
   f_{\pm}(\rho) = 1 - \big( \frac{\rho_\pm}{\rho} \big)^{D-2} \,.
  \end{align}
Physical branes without a naked singularity have more mass than charge, which corresponds
to  $\rho_+ > \rho_-$ or $k>0$. For this type of non-extremal deformation, the dilaton
$\hat{\phi}$ is proportional to $A$ and $B$, which are linearly related since $X=0$.
 \item {\bf Type 2 non-extremal $p$-branes:}\ $X \neq 0$ and $f = 0$.

These are the non-extremal black branes of \cite{Lu:1996hh}. The
deformation function $X$ reads
 \begin{align}
  e^{X} = 1 - \frac{k}{r^{2(D-2)}} \,,
 \end{align}
where $k$ is the deformation parameter. The absence of naked singularities requires $k$
to be positive. In this case, the dilaton $\hat{\phi}$ is not proportional to $A$ or $B$,
which are not linearly related.

The non-extremal D-instanton solutions \eqref{SolEq} fit exactly in this chain of
non-extremal $p$-branes for $p=-1$. Although the type 2 non-extremal $p$-branes are
defined in Minkowski space, we find that one can extend the formulae of \cite{Lu:1996hh}
to $p=-1$ branes in Euclidean space, i.e.~generalised D-instantons, by taking $f = 0$ and
$B \neq 0$.
  \end{itemize}
Both types of non-extremal $p$-branes break supersymmetry. A
special case is $p=0$, for which the regular type 1 and type 2
non-extremal $0$-branes are equivalent up to a coordinate
transformation in $r$. From the form of the metric \eqref{branes},
which has different world-volume isometries for $f = 0$ and $f
\neq 0$, it is clear that this is not the case for $p > 0$.

To relate the (multi-dilaton) instanton solutions of section 3 to the non-extremal
$p$-branes, it is instructive to reduce the $p$-branes over their $(p+1)$-dimensional
world-volume, including time. In complete analogy to the reduction over time of section
4.1, this will give rise to $p+1$ dilatons from the world-volume of the $p$-brane.
However, these are not all unrelated: for one thing, the dilatons corresponding to the
spatial world-volume will be proportional to each other, and can therefore be truncated
to a single dilaton. We will denote the dilaton from the spatial metric components
by $\varphi$, while the time-like component of the metric gives rise to $\tilde \varphi$.
In general, the reduction of non-extremal $p$-branes will therefore give rise to a
multi-instanton solution with three different dilatons, including the explicit dilaton
$\phi$:
 \begin{align}
  \hat{g}_{tt} \rightarrow \tilde \varphi \,, \qquad
  \hat{g}_{xx} \rightarrow \varphi \,, \qquad
  \hat{\phi} \rightarrow \phi \,.
 \end{align}
For the two types of non-extremal deformations considered here, however, there is always
a relation between the three dilatons, allowing a truncation to two
dilatons\footnote{This seems to indicate a generalisation of the non-extremal
deformations with both $X \neq 0$ and $f \neq 0$, reducing to a three-dilaton
instanton.}. For the type 1 deformations the dilatons $\phi$ and $\varphi$ are related,
as can be seen from the metric with $X=0$. Similarly, the type 2 deformations yield a
relation between $\varphi$ and $\tilde \varphi$ since $f=0$. Therefore, these
non-extremal $p$-branes reduce to multi-dilaton instanton solutions with two inequivalent
dilatons. Conversely, two-dilaton instanton solutions can uplift to either types of
non-extremal $p$-branes, by embedding these dilatons in different ways in the
higher-dimensional metric and dilaton.

It is interesting to investigate when these two dilatons can be related or reduce to one,
therefore corresponding to our explicit $SL(2,\mathbb{R})$ instanton solution
(\ref{preSolEq}) with only one dilaton. For the type 1 deformations, this is only
possible for the special case with $p=0$ and $a=0$. For these values, the dilatons $\phi$
and $\varphi$ vanish, leaving one with only $\tilde \varphi$. The constraint on $a$
implies $bc = 2$ which, as discussed in section 3, gives rise to the Reissner-Nordstr\"om
black hole.

For the type 2 deformations there are more possibilities to eliminate the dilaton $\phi$.
It can be achieved by requiring $a=0$, as we did for the uplift to black holes. For
general $p$, this leads to the following constraint on $b$:
\begin{equation}
  b = \sqrt{\frac{2(p+1)(D-2)}{D+p-1}} \,. \label{b-nondilaton}
\end{equation}
Note that this yields $bc=2$ for black holes with $p=0$. For these values of $b$, the
instanton solution (\ref{preSolEq}) can be uplifted to regular non-extremal non-dilatonic
$p$-branes. For higher values of $b$, the instanton solution uplifts to singular
non-extremal dilatonic $p$-branes. For these solutions to become regular, one must take
either $\mathsf{q}^2 \rightarrow 0$ or $q_- \rightarrow 0$, exactly like we found in the
$b c > 2$ discussion of section 4.3.

The uplift of the $SL(2,\mathbb{R})$ instanton solution (\ref{preSolEq}) to $p$-branes is
therefore very similar to the uplift to black holes. There is one value of $b$
\eqref{b-nondilaton} for which the instanton solution can be uplifted to a regular
non-extremal non-dilatonic $p$-brane of type 2. For higher values of $b$ one can obtain
singular non-extremal dilatonic $p$-branes of type 2, which only become regular on either
of the limits $\mathsf{q}^2 \rightarrow 0$ and $q_- \rightarrow 0$. By adding an extra
dilaton to the instanton solution one can also connect to the regular type 1 and type 2
non-extremal dilatonic $p$-branes.

\section{Instantons}\label{positive}

In the previous section we focused on the bulk behavior of the
three conjugacy classes of instanton-like solutions. In this
section we will investigate which of these solutions can be
interpreted as instantons. Instantons are defined to be solutions
to the Euclidean equations of motion with finite, non-zero value
of the action. They sometimes have a tunneling interpretation, but
more generically, they contribute to certain correlation functions
in the path integral, with terms that are exponentially suppressed
by the instanton action. These correlation functions then induce
new interactions in the effective action, and for the extremal,
1/2 BPS, D-instantons in type IIB in $D=10$, these effects are
captured by certain $SL(2,\bb{Z})$ modular functions that multiply
higher derivative terms like $R^4$ and their superpartners
\cite{Green:1997tv}. Before we study correlation functions and
effective interactions induced by non-extremal D-instantons, we
must first discuss the properties and show the finiteness of the
non-extremal instanton action. We will do this in such a way that
the special case of extremal D-instantons can easily be recovered.

\subsection{Instanton Action}\label{instantonaction}

The first thing we notice is that the action \eqref{EuclideanAction2}, evaluated on {\it
any} solution of \eqref{FieldEqs} vanishes. What is also bothersome about the Euclidean
action \eqref{EuclideanAction2} is that it is not bounded from below, not even in the
scalar sector. Such actions cannot be used for a semiclassical approximation in the path
integral, since fluctuations around the instanton will diverge. This problem can be
solved by adding boundary terms that guarantee a positive definite action for the
scalars, yielding at the same time a nonzero value for the instanton action.
These boundary terms can be understood as coming from dualising the magnetic
nine-form into the axion field $\chi$ \cite{Gibbons:1996vg,Green:1997tv}, subject to appropriate boundary conditions for the fields and their variations.
This dual formulation has a manifestly positive definite action (apart from the usual problems with the Einstein-Hilbert term), in which it is easy to derive a Bogomol'nyi bound and therefore, the semiclassical approximation is justified.
This procedure was also demonstrated in lower dimensions in \cite{Theis:2002er}.

The action for the dilaton and the ``magnetic'' $(D-1)$-form, dual
to the axion $\chi$, can be written as \cite{Green:1997tv}
\begin{equation}\label{dual-action}
\mathcal{S}_E = \int_\mathcal{M} \tfrac{1}{2} [ ( d\phi \wedge
\ast d\phi + e^{-b\phi} F_{D-1} \wedge \ast F_{D-1})]\,,
\end{equation}
supplemented by the constraint that $F_{D-1}$ is closed. Notice that in this
formalism, the $SL(2,\mathbb{R})$ symmetry is not manifest.

We can dualise back to the dilaton-axion system by introducing a Lagrange multiplier
$\chi$ that enforces the Bianchi-identity for $F_{D-1}$, i.e. ${\mathcal L}_{mult}=-\chi
dF_{D-1}$. If we now algebraically eliminate $F_{D-1}$ from the action by treating it as
a fundamental field (as opposed to treating it as a field strength) and using its
equation of motion,
\begin{equation}\label{F-chi}
F_{D-1}=-e^{b\phi}\ast d\chi\ ,
\end{equation}
we obtain the Euclidean action (\ref{EuclideanAction2}) in terms
of the axion $\chi$ plus the boundary term mentioned above.

It is now easy to show that this action satisfies a Bogomol'nyi bound
\cite{Green:1997tv}. Using the fact that, in a  Euclidean space, $\ast\ast A_p =
(-)^{(D-1)p}A_p$, where $A_p$ is a $p$-form, we can rewrite the action as follows:
\begin{equation}\label{Bog-action}
\mathcal{S}_E = \int_\mathcal{M} \tfrac{1}{2} [(d\phi \pm
e^{-b\phi /2} \ast F_{D-1}) \wedge \ast (d\phi \pm e^{-b\phi /2}
\ast F_{D-1}) \mp (-)^{D}\tfrac{4}{b}d(e^{-b\phi /2} F_{D-1})]\,,
\end{equation}
where we have used the fact that $dF_{D-1}=0$. Since the first
term is positive semi-definite $\mathcal{S}_E$ is bounded from
below by a topological surface term given by the last term in
\eqref{Bog-action}. The bound is saturated when the Bogomol'nyi
equation
\begin{equation}\label{Bog-eqn}
\ast F_{D-1}= \mp e^{b\phi/2}d\phi\ ,
\end{equation}
is satisfied. The $\mp$ distinguishes instantons from
anti-instantons, and for simplicity, we will use the upper sign
from now on. Using \eqref{F-chi}, one can write the Bogomol'nyi
equation as
\begin{equation}
d\chi = -e^{-b\phi/2}d\phi\,,
\end{equation}
and one can check explicitly that the instanton solutions with $\mathsf{q}^2=0$, given in
\eqref{instlimEq}, satisfy this bound. They are therefore rightfully called extremal. The
instanton action can then easily be evaluated, and has only a contribution from the
boundary at infinity,
\begin{equation}\label{extr-inst-act-infty}
\mathcal{S}^{\infty}_{{inst}}=\frac{4}{b^2} (D-2)
\mathcal{V}ol(S^{D-1})\frac{|bcq_-|}{g_s^{b/2}}\ ,
\end{equation}
while the contribution from $r=0$ vanishes.

For $D=10$ and $b=2$, this value of the instanton action precisely 
coincides with \cite{Gibbons:1996vg}. For other values of $b$, we 
notice the dependence of $g_s$ on $b$. In ten dimensions, the only 
possible value for $b$ compatible with maximal supersymmetry is 
$b=2$. One then finds that the instanton action depends linearly on 
the inverse string coupling constant. In lower dimensions this is 
not necessarily so, and more values for $b$ are possible, depending 
on whether $\chi$ comes from the RR sector or from the NS sector. 
This would imply different kinds of instanton effects, with 
instanton actions that depend on different powers of the string 
coupling constant. This indeed happens for instance in four 
dimensions, after compactifying type IIA strings on a Calabi-Yau 
threefold. There are D-instantons coming from wrapping (Euclidean) 
D2 branes around a supersymmetric three-cycle, and there are 
NS5-brane instantons coming from wrapping the NS5-brane around the 
entire Calabi-Yau. As explained in \cite{Becker:1995kb}, such 
instanton effects are weighted with different powers of $g_s$ in the 
instanton action. This was also explicitly demonstrated in 
\cite{Theis:2002er, Davidse:2003ww, Davidse:2004gg}. In our 
notation, they correspond\footnote{This corrects a minor mistake in 
the previous version and in the version published in {\it JHEP}. In 
our conventions, the $D=4$ dilaton is related to the $D=10$ string 
dilaton by a factor of 2, see \cite{proc} for further details and 
implications of this correction.} to $b=1$ and $b=2$. Our results in 
\eqref{extr-inst-act-infty} are consistent with these observations.

Notice also that the instanton action is proportional to $q_-$. For extremal instantons,
this is precisely the mass of the corresponding black hole in one dimension higher, see
\eqref{RN-instanton-relation}. This is a generic characteristic of the instanton-soliton
correspondence that we know from field theories. There, the Euclidean action in $D$
dimensions equals the Hamiltonian in $D+1$ dimensions, and the instanton action equals
the soliton mass. It is interesting to see that this also happens for theories with
gravity.

We now turn to the case of non-extremal instantons, and focus first
on the case of $\mathsf{q}^2> 0$. The solutions
\eqref{SolEq} for the dilaton and axion fields can be written as
\begin{equation}
d\phi = \frac{2}{b} \coth(H+C_1) dH\ ,\qquad
e^{-b\phi/2}F_{D-1}=\frac{2}{b}\frac{\ast dH}{\sinh(H+C_1)}\,,
\end{equation}
and do not satisfy the Bogomol'nyi equation \eqref{Bog-eqn}. To evaluate the action on
this non-extremal instanton solution, we plug in these expressions into the bulk action
\eqref{dual-action}, and find
\begin{equation}
\mathcal{S}_{inst}=\frac{2}{b^2}\int\,d\Big(\{H-2 \coth(H+C_1)\}
\ast dH\Big)\,,\label{inst-act}
\end{equation}
which is again a total derivative term. Using Stokes theorem, we therefore only pick up
contributions from the boundaries. Since the $\mathsf{q}^2> 0$ instantons have a
curvature singularity at $r=r_c$ (see section 3.1), one can take these boundaries at
$r=\infty$ and at $r=r_c$. In terms of the variable $H$, this corresponds to $H=0$ and
$H=\infty$ respectively\footnote{Without loss of generality, we can choose
$\mathsf{q}>0$.}. We remind again that we have taken $C_1$ to be positive, in order to
avoid further singularities in the scalar sector when $H+C_1=0$.

Evaluating the Einstein-Hilbert term on the solution in \eqref{SolEq} we find
the following:
\begin{align}
- \int_{\mathcal{M}}R &= -\frac{2}{b^2}\,\int_{\mathcal{M}}d(H
\ast dH)\,,
\end{align}
which precisely cancels the first term of the scalar action \eqref{inst-act}.
Strictly speaking, both these terms diverge at the boundary $r=r_c$ as one can show, and
need to be regularized. For the Einstein-Hilbert term, this needs to be done
in combination with the Gibbons-Hawking term \cite{Gibbons:1977ue},
\begin{align} \label{EH-act}
  \mathcal{S}_{EH} &= - \Big(\int_{\mathcal{M}}R + 2\int_{\partial{\mathcal{M}}}(K-K_0)\Big)\,,
\end{align}
where $\mathcal{M}$ is the $D$-dimensional Euclidean space and $\partial{\mathcal{M}}$ is
the boundary. In the second term, $K$ is the trace of the extrinsic curvature of the
boundary and $K_0$ the extrinsic curvature one would find for flat space, which is
subtracted to normalise the value of the action. One can check that from the boundary at
$r=\infty$, there is no contribution to \eqref{EH-act}. At $r=r_c$, the metric has a
curvature singularity and the dilaton blows up; therefore the supergravity approximation
breaks down. One would have to rely on higher order string theory corrections to
regularize the contribution from $r_c$. Whatever the precise contribution is, we remark
that the gravitational action \eqref{EH-act} is an $SL(2,\mathbb{R})$ invariant,
independent of the string coupling constant. It can therefore only be a function of
$\mathsf{q}^2$.

If we assume that the regularization is such that there is still a cancellation with the
first term in \eqref{inst-act}, we only have contributions coming from the second terms
of both integrals \eqref{inst-act} and \eqref{EH-act}. We first discuss the boundary at
$r = \infty$. The contribution from \eqref{EH-act} vanishes, while \eqref{inst-act}
yields a contribution
\begin{align} \label{nonextr-inst-act-infty}
 \mathcal{S}^{\infty}_{inst} & = \frac{4}{b^2}\,(D-2) \mathcal{V}ol(S^{D-1})\,{b\,c}\,
\Big(\mathsf{q}\, \coth C_1 \Big) \,, \notag \\
 & = \frac{4}{b^2}\,(D-2)\, \mathcal{V}ol(S^{D-1})\,b\,c\,\Big(\sqrt{\mathsf{q}^2+\frac{q_-^2}{g_s^b}} \Big)\ .
\end{align}
In the second line, we have used the relation between $C_1$ and the asymptotic value of
the dilaton, $g_s^{b}=(q_-/\mathsf{q})^2\,\sinh^2 C_1$.

For $\mathsf{q}^2=0$, \eqref{nonextr-inst-act-infty} precisely yields back the result for
the extremal instanton, see \eqref{extr-inst-act-infty}. There we made the relation
between the instanton action and the black hole mass in one dimension higher. Also for
the non-extremal instanton, such a relation seems to holds. Indeed, from the mass formula
for the non-extremal black hole in terms of the instanton parameters, one has that
$\mathsf{q}\,\coth C_1=\sqrt{\mathsf{q}^2 +q_-^2}$, and the string coupling constant is
set to unity. One therefore sees that the contribution to the instanton action from the
boundary at infinity is proportional to the black hole mass in one dimension higher.

The boundary at $r=r_c$ receives contributions from both integrals \eqref{inst-act} and
\eqref{EH-act}, which add up to
\begin{align} \label{nonextr-inst-act-rc}
 \mathcal{S}^{r_c}_{inst} & = \frac{4}{b^2}\,(D-2) \mathcal{V}ol(S^{D-1})\,{b\,c}\,
\Big(\mathsf{q}\,(\frac{bc}{2} -1) \Big) \,.
\end{align}
Note that this contribution vanishes for the case $bc=2$, while it is positive for
$bc>2$. However, as discussed above, it is not at all clear whether this contribution to
the integrals \eqref{inst-act} and \eqref{EH-act} should be included in the instanton
action, since it is calculated in a region of space where the supergravity approximation
is no longer valid. It might well be that string corrections smooth out the singularity
at $r=r_c$, leaving one with only the contribution \eqref{nonextr-inst-act-infty} from
$r=r_\infty$.

We now turn to the case of $\mathsf{q}^2 < 0$, or with $\mathsf{q}=i \mathsf{\tilde q}$,
a positive $\mathsf{\tilde q}^2 >0$. A similar calculation as for $\mathsf{q}^2>0$ shows
that, for the solution \eqref{qminussol}, we have
\begin{equation}
d\phi = \frac{2}{b} \cot(\tilde{H}+\tilde{C_1}) d\tilde{H}\ ,\qquad
e^{-b\phi/2}F_{D-1}=\frac{2}{b}\frac{\ast d\tilde{H}}{\sin(\tilde{H}
+\tilde{C_1})}\,,
\end{equation}
where
\begin{equation}
\tilde{H}=bc \arctan(\frac{\tilde{q}}{r^{D-2}})\,,
\end{equation}
is a harmonic function over the geometry given by the metric in \eqref{qminussol}.
Plugging in these expressions into the bulk action \eqref{dual-action}, we find
\begin{equation}
\mathcal{S}_{inst}=-\frac{2}{b^2}\int\,d\Big(\{\tilde{H}+2 \cot(\tilde{H} +{\tilde
C_1})\} \ast d\tilde{H}\Big)\,.\label{qtilde-inst-act}
\end{equation}
Since this is a total derivative, we can use Stokes theorem again to reduce it to an
integral over the boundaries. These boundaries are at $r=\infty $ and $r=0$, where we
required that $bc < 2$, as discussed in section 3.1. In contrast to the discussion of the
$r=r_c$ boundary for $\mathsf{q}^2 >0$, the instanton solution is perfectly regular
everywhere, in particular at both boundaries. Therefore the contribution from the
boundary at $r=0$ can also be trusted.

In addition to the above action, one also needs to include the gravitational contribution
\eqref{EH-act}. Similar to the case of $\mathsf{q}^2 > 0$, the first term of
\eqref{qtilde-inst-act} is cancelled by the contribution from the Ricci scalar. We
anticipate the Gibbons-Hawking term not to contribute, since the two asymptotic
geometries at $r=0$ and $r=\infty$ are equivalent due to the $\mathbb{Z}_2$-symmetry
\eqref{Z2-isometry} and therefore their contributions should cancel.

Therefore the $\mathsf{q}^2 <0$ instanton action has contributions only from the second
term of \eqref{qtilde-inst-act} from both boundaries at $r=0$ and $r=\infty$:
\begin{align} \label{tildeq-inst-act}
 \mathcal{S}_{inst}^\infty & = \frac{4}{b^2}\,(D-2) \mathcal{V}ol(S^{D-1})\,{b\,c}\,
\mathsf{\tilde {q}}\,\Big(\cot \tilde{C}_1 \Big)
\,, \notag \\
\mathcal{S}_{inst}^0 & = \frac{4}{b^2}\,(D-2) \mathcal{V}ol(S^{D-1})\,{b\,c}\,
\mathsf{\tilde {q}}\,\Big(-\cot (\tilde{C}_1 + bc \frac{\pi}{2})\Big) \,.
\end{align}
Due to the fact that $\tilde{C}_1$ and $\tilde{C}_1 + bc\pi/2$ are on the same branch of
the cotangent (due to the restriction of regular scalars for $0<r<\infty$, which can only
be achieved for $bc<2$, see section~3.1), the total instanton action is manifestly positive
definite. In the neighborhood of $bc\approx 2$, the instanton action becomes very large,
and the limit to the extremal point where $bc=2$, is discontinuous. This shows that this
instanton is completely disconnected from the extremal D-instanton.

Using the
asymptotic value of the dilaton in \eqref{qminussol}, we have $g_s^{b} =
(q_-/\mathsf{\tilde q})^2\sin^2 \tilde {C}_1$, and therefore $\mathsf{\tilde q}^2 <
q_-^2/g_s^b$. Assuming that $\cot \tilde {C}_1>0$, the contribution from infinity is positive and can be rewritten as
\begin{align}
\mathcal{S}_{inst}^{\infty}=
\frac{4}{b^2}\,(D-2)\, \mathcal{V}ol(S^{D-1})\,b\,c\,\sqrt{\frac{q_-^2}{g_s^b}
-\mathsf{\tilde q}^2}\ ,
\end{align}
which is the analytic continuation of the result with $\mathsf{q}^2>0$.

\subsection{Correlation Functions}

Once the instanton solutions are established, one studies their effect in the path
integral. As for D-instantons in ten-dimensional IIB, they contribute to certain
correlation functions via the insertion of fermionic zero modes. For the D-instanton,
which is 1/2 BPS, there are sixteen fermionic zero modes. These are solutions for the
fluctuations that satisfy the linearised Dirac equation in the presence of the instanton.
All of these zero modes can be generated by acting with the broken supersymmetries on the
purely bosonic instanton solution. For the non-extremal instantons, no supersymmetries
are preserved, so there are more fermionic zero modes. Let us focus for simplicity on
ten-dimensional type IIB. Since all the supercharges are broken, one can generate 32
fermionic zero modes. The path integral measure contains an integration over these
fermionic collective coordinates, and to have a non-vanishing result, one must therefore
insert 32 dilatinos in the path integral. Based on this counting argument of fermionic
zero modes, a 32-point correlator of dilatinos would be non-zero, and induce new terms in
the effective action, containing 32 dilatinos. In the full effective action, such terms
are related to higher curvature terms like e.g. certain contractions of $R^8$. An
explicit instanton calculation should be done to determine the non-perturbative
contribution to the function that multiplies $R^8$. As for the D-instanton, we expect
that the contributions of the instantons with different $\mathsf{q}^2$-values build up a
modular form with respect to $SL(2,\bb{Z})$, possibly after integrating over
$\mathsf{q}^2$.

These issues, though important, lie beyond the scope of this
paper, and are left for further investigation.

\section{Discussion}\label{conclusions}

In this paper we investigated non-extremal instantons in string theory that are solutions
of a gravity-dilaton-axion system with dilaton coupling parameter $b$. In particular, we
constructed an $SL(2,\mathbb{R})$ family of radially symmetric instanton-like solutions
in all conjugacy classes labelled by $\mathsf{q}^2$. Among these is the
(anti-)D-instanton solution with $\mathsf{q}^2 = 0$. For special values of the dilaton
coupling parameter this solution is half-supersymmetric. The instanton solutions in the
other two conjugacy classes, with $\mathsf{q}^2 > 0$ and $\mathsf{q}^2 < 0$, are
non-supersymmetric and can be viewed as the non-extremal version of the
(anti-)D-instanton. This view is confirmed by the property that instantons in these two
conjugacy classes, for $bc \ge 2$ with $c$ defined in \eqref{ceq}, can be uplifted to
non-extremal black holes.

We stressed the wormhole nature of the instanton solutions. We found that each conjugacy
class leads to a wormhole geometry provided the corresponding instanton is given in a
particular metric frame:
\begin{eqnarray}
 \mathsf{q}^2 > 0 &\leftrightarrow& {\rm dual\ frame\ metric\ (only~for~} bc=2 {\rm ~or~} q_-=0) \nonumber\\
 \mathsf{q}^2 = 0  &\leftrightarrow&  {\rm string\ frame\ metric}    \\
 \mathsf{q}^2 < 0  &\leftrightarrow& {\rm Einstein\ frame\ metric}\nonumber
\end{eqnarray}
For all these case the metric takes the form \eqref{genwormholemetric}, with the specific
values given in section 3.2.

Not all instanton solutions we constructed are regular and not all can be uplifted to
black holes. The non-extremal instantons in the $\mathsf{q}^2 > 0$ conjugacy class all
have a curvature singularity at $r=r_{\rm c}$, see \eqref{rcritical}. Only the $bc=2$
instanton can be uplifted to a regular non-extremal RN black hole with the singularity
being resolved as a coordinate singularity at the outer event horizon of the RN black
hole. The singularity remains for $bc>2$ and in that case can be resolved by adding an
extra dilaton to the original system \cite{Cremmer:1998em}. Two exceptions are the limits
$\mathsf{q}^2 \rightarrow 0$ or $q_- \rightarrow 0$, which correspond to the extremal and
Schwarzschild black hole solutions, respectively. Finally, the instantons in the
$\mathsf{q}^2 < 0$ conjugacy class are only regular for $bc <2$. These instantons can
never be uplifted to black holes.

We have also considered the uplift of our instanton solutions to $p$-branes. It turns out
that an instanton can only be uplifted over a $(p+1)$-torus to a $p$-brane provided the
dilaton coupling satisfies (following from \eqref{b-nondilaton})
 \begin{align}
   b \, c \geq \sqrt{\frac{4(p+1)(D-1)}{D+p-1}} \,. \label{b-bound}
 \end{align}
For the case that saturates this bound, the instanton with $\mathsf{q}^2 \geq 0$ uplifts
to a regular non-dilatonic $p$-brane. For larger values of $b$, the instanton solution
\eqref{preSolEq} with $\mathsf{q}^2 > 0$ uplifts to a singular limit of the dilatonic
$p$-branes of \cite{Lu:1996hh}. These solutions only become regular in the limit
$\mathsf{q}^2 \rightarrow 0$ or $q_- \rightarrow 0$. A summary of the possible regular
solutions is given in table \ref{regular}. Alternatively, we have discussed the
possibility of adding an extra dilaton to the instanton solution \cite{Cremmer:1998em},
which allows for the uplift to the regular dilatonic $p$-branes of both type 1 and type
2.

\begin{table}[ht]
\begin{center}
\hspace{-1cm}
\begin{tabular}{||c||c|l||}
\hline \rule[-1mm]{0mm}{6mm}
 $bc$ & Dimension & Regular solutions \\
\hline \hline \rule[-1mm]{0mm}{6mm}
 $<2$ & $D$ & Instantons with $\mathsf{q}^2 \leq 0$, see \eqref{qminussol} \\
\hline\rule[-1mm]{0mm}{6mm}
 $=2$ & $D+1$ & RN black holes with $\mathsf{q}^2 \geq 0$, see \eqref{RN}, or \\
 & & Schwarzschild black holes with $\mathsf{q}^2 > 0$, $q_- = 0$ \\
\hline \rule[-1mm]{0mm}{6mm}
 $>2$ & $D+1$ & Dilatonic black holes with $\mathsf{q}^2 = 0$ or \\
 & & Schwarzschild black holes with $\mathsf{q}^2 > 0$, $q_- = 0$ \\
\hline\rule[-1mm]{0mm}{6mm}
 $=$ in \eqref{b-bound} & $D+p+1$ & Non-dilatonic $p$-branes with $\mathsf{q}^2 \geq 0$ \\
\hline\rule[-1mm]{0mm}{6mm}
 $>$ in \eqref{b-bound} & $D+p+1$ & Dilatonic $p$-branes with $\mathsf{q}^2 = 0$ or\\
&&  $\mathsf{q}^2 > 0$, $q_- = 0$ \\
\hline
\end{tabular}
\caption{\it The regular instanton, black hole and $p$-brane solutions that are obtained,
depending on the dilaton coupling parameter $b$, the conjugacy class $\mathsf{q}^2$ and
the charge $q_-$.} \label{regular}
\end{center}
\end{table}

For the particular value $b=2$, corresponding to $\Delta = 4$, there is another
higher-dimensional origin. In this special case, the $D$-dimensional extremal instanton
can be uplifted to a gravitational wave in $D+2$ dimensions \cite{Tseytlin:1997ne}.
Similarly, the other two conjugacy classes uplift to purely gravitational solutions in
$D+2$ dimensions which we denominate ``non-extremal waves''. The terminology is slightly
misleading since the uplift only leads to a time-independent solution. Whether this
solution can be extended to a time-dependent wave-like solution remains to be seen. It is
also interesting to note the following curiosity. The source term for a pp-wave is a
massless particle, i.e.~a particle with a null-momentum vector: $p^2=0$. It is suggestive
to associate the source terms for the other two conjugacy classes with massive particles
($p^2>0$) and tachyonic particles ($p^2<0$). We leave this for a future investigation.

In the second part of this paper we investigated the possibility whether the non-extremal
instantons might contribute to certain correlation functions in string theory. For this
application it is a prerequisite that there is a well-defined and finite instanton
action. Mimicking the calculation of the standard D-instanton action we found that for
$\mathsf{q}^2 > 0$ the contribution from infinity to the instanton action, for all values
of $b$, is given by the elegant formula \eqref{nonextr-inst-act-infty}. This action
reduces to the standard D-instanton action for $\mathsf{q}^2 = 0$. Having a finite
action, the non-extremal instantons might contribute to certain correlation functions. In
the case of type IIB string theory we conjectured that non-extremal instantons contribute
to the $R^8$ terms in the string effective action in the same way that the extremal
D-instantons contribute to the $R^4$ terms in the same action. Whether the fact that all
supersymmetries are broken by the non-extremal instantons poses problems remains to be
seen. An explicit instanton calculation should decide whether our conjecture is correct.
This and related issues we leave for future investigation.

\section*{Acknowledgements}

We thank Pierre van Baal, Martin Cederwall, Ludde Edgren, Rom\'an Linares, Tom\'as
Ort\'\i n, Andr\'e Ploegh, Bert Schellekens and Tim de Wit for useful discussions. This
work is supported in part by the Spanish grant BFM2003-01090 and the European Community's
Human Potential Programme under contract HPRN-CT-2000-00131 Quantum Spacetime, in which
E.B. and D.R.~are associated to Utrecht University. The work of U.G.~is funded by the
Swedish Research Council.


\begin{thebibliography}{10}
\bibliographystyle{utphysmodb}


\bibitem{Bergshoeff:2002mb}
E.~Bergshoeff, U.~Gran and D.~Roest,  {\em Type IIB seven-brane solutions from
  nine-dimensional domain walls}, Class. Quant. Grav. {\bf 19} (2002)
  4207--4226
{{\tt hep-th/0203202}}.

\bibitem{Bergshoeff:1996ui}
E.~Bergshoeff, M.~de~Roo, M.~B. Green, G.~Papadopoulos and P.~K. Townsend,
  {\em Duality of Type II 7-branes and 8-branes}, Nucl. Phys. {\bf B470} (1996)
  113--135
{{\tt [hep-th/9601150]}}.

\bibitem{Bergshoeff:2004nq}
E.~Bergshoeff, M.~Nielsen and D.~Roest,  {\em The domain walls of gauged
  maximal supergravities and their M-theory origin},
{{\tt hep-th/0404100}}.

\bibitem{Gibbons:1996vg}
G.~W. Gibbons, M.~B. Green and M.~J. Perry,  {\em Instantons and Seven-Branes
  in Type IIB Superstring Theory}, Phys. Lett. {\bf B370} (1996) 37--44
{{\tt hep-th/9511080}}.

\bibitem{Green:1997tv}
M.~B. Green and M.~Gutperle,  {\em Effects of D-instantons}, Nucl. Phys. {\bf
  B498} (1997) 195--227
{{\tt hep-th/9701093}}.

\bibitem{Giddings:1989bq}
S.~B. Giddings and A.~Strominger,  {\em String Wormholes}, Phys. Lett. {\bf
  B230} (1989)
46.

\bibitem{Coule:1989xu}
D.~H.~Coule and K.~i.~Maeda,
Class.\ Quant.\ Grav.\  {\bf 7} (1990) 955.

\bibitem{Tseytlin:1997ne}
A.~A. Tseytlin,  {\em Type IIB instanton as a wave in twelve dimensions}, Phys.
  Rev. Lett. {\bf 78} (1997) 1864--1867
{{\tt hep-th/9612164}}.

\bibitem{Kim:1997hq}
J.~Y. Kim, H.~W. Lee and Y.~S. Myung,  {\em D-instanton and D-wormhole}, Phys.
  Lett. {\bf B400} (1997) 32--36
{{\tt hep-th/9612249}}.

\bibitem{Einhorn:2000ct}
M.~B. Einhorn and L.~A. Pando~Zayas,  {\em On seven-brane and instanton
  solutions of type IIB}, Nucl. Phys. {\bf B582} (2000) 216--230
{{\tt hep-th/0003072}}.

\bibitem{Einhorn:2002am}
M.~B. Einhorn,  {\em Instantons of type IIB supergravity in ten dimensions},
  Phys. Rev. {\bf D66} (2002) 105026
{{\tt hep-th/0201244}}.

\bibitem{Gutperle:2002km}
M.~Gutperle and W.~Sabra,  {\em Instantons and wormholes in Minkowski and (A)dS
  spaces}, Nucl. Phys. {\bf B647} (2002) 344--356
{{\tt hep-th/0206153}}.

\bibitem{Kim:2003js}
J.~Y.~Kim, Y.~b.~Kim and J.~E.~Hetrick,
arXiv:hep-th/0301191.

\bibitem{Khoze:2003yk}
V.~V. Khoze,  {\em From branes to branes},
{{\tt hep-th/0311065}}.

\bibitem{Bergshoeff:1998ry}
E.~Bergshoeff and K.~Behrndt,  {\em D-instantons and asymptotic geometries},
  Class. Quant. Grav. {\bf 15} (1998) 1801--1813
{{\tt hep-th/9803090}}.

\bibitem{Bergshoeff:2000qu}
E.~Bergshoeff and A.~Van~Proeyen,  {\em The many faces of $\:OSp\,(1|\:32)$},
  Class. Quant. Grav. {\bf 17} (2000) 3277--3304
{{\tt hep-th/0003261}}.

\bibitem{Einhorn:2002sj}
M.~B. Einhorn,  {\em Instantons and SL(2,R) symmetry in type IIB supergravity},
{{\tt hep-th/0212322}}.

\bibitem{Meessen:1998qm}
P.~Meessen and T.~Ort\'\i{}n,  {\em An Sl(2,Z) multiplet of nine-dimensional
  type II supergravity theories}, Nucl. Phys. {\bf B541} (1999) 195--245
{{\tt hep-th/9806120}}.

\bibitem{Boonstra:1998mp}
H.~J. Boonstra, K.~Skenderis and P.~K. Townsend,  {\em The domain wall/QFT
  correspondence}, JHEP {\bf 01} (1999) 003
{{\tt hep-th/9807137}}.

\bibitem{Behrndt:1999mk}
K.~Behrndt, E.~Bergshoeff, R.~Halbersma and J.~P. van~der Schaar,  {\em On
  domain-wall/QFT dualities in various dimensions}, Class. Quant. Grav. {\bf
  16} (1999) 3517--3552
{{\tt hep-th/9907006}}.

\bibitem{Cremmer:1998em}
E.~Cremmer, I.~V. Lavrinenko, H.~Lu, C.~N. Pope, K.~S. Stelle and T.~A. Tran,
  {\em Euclidean-signature supergravities, dualities and instantons}, Nucl.
  Phys. {\bf B534} (1998) 40--82
{{\tt hep-th/9803259}}.

\bibitem{Lu:1995cs}
H.~Lu, C.~N. Pope, E.~Sezgin and K.~S. Stelle,  {\em Stainless super p-branes},
  Nucl. Phys. {\bf B456} (1995) 669--698
{{\tt hep-th/9508042}}.

\bibitem{Townsend:1997ku}
P.~K. Townsend,  {\em Black holes},
{{\tt gr-qc/9707012}}.

\bibitem{Lu:1996hh}
H.~Lu, C.~N. Pope and K.~W. Xu,  {\em Liouville and Toda Solutions of
  M-theory}, Mod. Phys. Lett. {\bf A11} (1996) 1785--1796
{{\tt hep-th/9604058}}.

\bibitem{Gibbons:1988ps}
G.~W. Gibbons and K.-i. Maeda,  {\em Black holes and membranes in higher
  dimensional theories with dilaton fields}, Nucl. Phys. {\bf B298} (1988)
741.

\bibitem{Horowitz:1991cd}
G.~T. Horowitz and A.~Strominger,  {\em Black strings and P-branes}, Nucl.
  Phys. {\bf B360} (1991)
197--209.

\bibitem{Duff:1994ye}
M.~J. Duff and J.~X. Lu,  {\em Black and super p-branes in diverse dimensions},
  Nucl. Phys. {\bf B416} (1994) 301--334
{{\tt hep-th/9306052}}.

\bibitem{Theis:2002er}
U.~Theis and S.~Vandoren,  {\em Instantons in the double-tensor multiplet},
  JHEP {\bf 09} (2002) 059
{{\tt hep-th/0208145}}.

\bibitem{Becker:1995kb}
K.~Becker, M.~Becker and A.~Strominger,  {\em Five-branes, membranes and
  nonperturbative string theory}, Nucl. Phys. {\bf B456} (1995) 130--152
{{\tt hep-th/9507158}}.

\bibitem{Davidse:2003ww}
M.~Davidse, M.~de~Vroome, U.~Theis and S.~Vandoren,  {\em Instanton solutions
  for the universal hypermultiplet}, Fortsch. Phys. {\bf 52} (2004) 696--701
{{\tt hep-th/0309220}}.

\bibitem{Davidse:2004gg}
M.~Davidse, U.~Theis and S.~Vandoren,  {\em Fivebrane instanton corrections to
  the universal hypermultiplet},
{{\tt hep-th/0404147}}.

\bibitem{proc}
E.~Bergoeshoeff, A.~Collinucci, U.~Gran, D.~Roest and S.~Vandoren, 
{\em Non-extremal instantons and wormholes in string theory}, 
proceedings of the RTN2004 workshop, 5-10 September, Kolymbari, 
Crete, Greece, {{\tt hep-th/0412183}}.

\bibitem{Gibbons:1977ue}
G.~W. Gibbons and S.~W. Hawking,  {\em Action integrals and partition functions
  in quantum gravity}, Phys. Rev. {\bf D15} (1977)
2752--2756.


\end{thebibliography}
\end{document}